\documentclass[aps,amsmath,amssymb,twocolumn,showpacs]{revtex4-1}

\usepackage{graphicx}
\usepackage{dcolumn}
\usepackage{bm}
\usepackage{color}

\begin{document}

\title{Thermal discrete dipole approximation for the description of thermal emission and radiative
heat transfer of magneto-optical systems}

\author{R. M. Abraham Ekeroth$^{1,2}$}
\author{A. Garc\'{\i}a-Mart\'{\i}n$^1$}
\author{J. C. Cuevas$^{3,4}$}
\email{juancarlos.cuevas@uam.es}

\affiliation{$^1$IMM-Instituto de Microelectr\'onica de Madrid (CNM-CSIC), Isaac Newton 8,
PTM, Tres Cantos, E-28760 Madrid, Spain}
\affiliation{$^{2}$Instituto de F\'{\i}sica Arroyo Seco, Universidad Nacional del Centro 
de la Provincia de Buenos Aires, Pinto 399, 7000 Tandil, Argentina}
\affiliation{$^3$Departamento de F\'{\i}sica Te\'orica de la Materia Condensada and Condensed Matter 
Physics Center (IFIMAC), Universidad Aut\'onoma de Madrid, E-28049 Madrid, Spain}
\affiliation{$^4$Department of Physics, University of Konstanz, D-78457 Konstanz, Germany}

\date{\today}

\begin{abstract}
We present here a generalization of the thermal discrete dipole approximation (TDDA) that allows 
us to describe the near-field radiative heat transfer between finite objects of arbitrary shape that
exhibit magneto-optical (MO) activity. We also extend the TDDA approach to describe the thermal 
emission of a finite object with and without MO activity. Our method is also valid for optically
anisotropic materials described by an arbitrary permittivity tensor and we provide simple closed 
formulas for the basic thermal quantities that considerably simplify the implementation of TDDA method. 
Moreover, we show that employing our TDDA approach one can rigorously demonstrate Kirchhoff's radiation 
law relating the emissivity and absorptivity of an arbitrary MO object. Our work paves the way for the 
theoretical study of the active control of emission and radiative heat transfer between MO systems of 
arbitrary size and shape.    
\end{abstract}

\maketitle

\section{Introduction}
There is presently a great interest in the study of the radiative heat transfer between closely 
placed objects \cite{Song2015b}. The reason for that is the experimental verification of the 
prediction that the radiative heat transfer between two bodies can be greatly enhanced when they are 
brought sufficiently close to each other \cite{Polder1971,Rytov1953}. This enhancement can lead to overcome 
the far-field limit set by the Stefan-Boltzmann law for black bodies and it takes place when the gap between 
the two objects is smaller than the thermal wavelength (9.6 $\mu$m at room temperature). Such an 
enhanced thermal radiation exchange originates from the contribution of evanescent waves that
dominate the near-field regime and by now it has been observed in numerous experiments 
\cite{Kittel2005,Rousseau2009,Shen2009,Ottens2011,Kralik2012,Zwol2012a,Zwol2012b,Guha2012,Worbes2013,
St-Gelais2014,Song2015a,Kim2015,St-Gelais2016,Song2016,Bernardi2016,Cui2017}. These experiments have in turn
triggered off the hope that near-field radiative heat transfer (NFRHT) could lead to new applications
in the context of thermal technologies such as thermophotovoltaics \cite{Laroche2006,Lenert2014,Bierman2016}, 
heat-assisted magnetic recording \cite{Challener2009,Stipe2010}, scanning thermal microscopy 
\cite{Wilde2006,Kittel2008,Jones2013}, nanolithography \cite{Pendry1999} or thermal management 
\cite{Otey2010,Ben-Abdallah2014}, just to mention a few. 

The theoretical description of NFRHT is usually done in the framework of fluctuational electrodynamics 
(FE) introduced by Rytov in the 1950s \cite{Rytov1953,Rytov1989}. In this framework, it is assumed that 
the thermal radiation is generated by random, thermally activated electric currents in the interior of 
a material. These currents vanish on average, but their correlations are given by the fluctuation-dissipation
theorem \cite{Landau1980,Keldysh1994,Joulain2005}. Thus, the technical problem involved in the description of 
the radiative heat transfer between two finite objects is to find the solution of the stochastic 
Maxwell equations with random electric currents as radiation sources. This problem can be quite
challenging and an analytical solution can only be obtained in a few cases with simple geometries like, for 
instance, two parallel plates \cite{Polder1971}, two spheres \cite{Narayanaswamy2008} or a sphere
in front of a plate \cite{Otey2011}. In general, in order to solve this problem for complex geometries,
which is usually necessary for a comparison with the experiments, one has to resort to numerical methods. 
In this respect, a lot of progress has been done in recent years and standard numerical methods in 
electromagnetism have already been combined with FE to describe NFRHT between objects of arbitrary 
size and shape (for a review see Ref.~\cite{Otey2014}). Those methods include, among others, the scattering 
matrix approach \cite{Kruger2011,Kruger2012}, finite-difference time- and frequency-domain methods 
\cite{Rodriguez2011,Liu2013,Datas2013,Wen2010}, the boundary element method \cite{Rodriguez2012,Rodriguez2013},
and volumen-integral-equation methods \cite{Polimeridis2015}. Recently Francoeur and coworkers adapted the 
discrete dipole approximation (DDA) to describe the NFRHT between two optically isotropic objects of arbitrary 
shape \cite{Edalatpour2014,Edalatpour2015,Edalatpour2016}. This new approach has been termed as thermal 
discrete dipole approximation (TDDA). It describes the NFRHT between two bodies in the framework
of FE by discretizing the objects in terms of point dipoles in the spirit of the DDA method that 
is widely used for describing the scattering and absorption of light by small particles 
\cite{Purcell1973,Draine1988,Draine1994,Yurkin2007}. The goal of this work is to generalize the 
TDDA method to describe the radiative heat transfer and the thermal emission in magneto-optical (MO) systems.

MO objects are of great interest in the context of NFRHT. Thus for instance, in a recent work
by some of us we showed that the NFRHT between two parallel plates made of doped semiconductors 
can be largely tuned by applying a static external magnetic field \cite{Moncada-Villa2015}.
Doped semiconductors under a magnetic field present a sizable MO activity, that can be
controlled by changing the magnitude and the direction of the field. More recently, it has been 
predicted that the NFRHT between several MO particles, also made of doped semiconductors, can 
lead to the appearance of striking phenomena such a near-field thermal Hall effect 
\cite{Ben-Abdallah2016} or the existence of a persistent directional heat current \cite{Zhu2016}. 

Motivated by these recent developments in the context of NFRHT of MO systems, we present here a 
generalization of the TDDA method to describe the radiative heat transfer between MO objects of 
arbitrary shape, something that is still missing in the literature. To be precise, we present a 
TDDA method that is valid for optically anisotropic systems that can be described by an arbitrary 
electric permittivity tensor (with $\mu=1$). We also extend the TDDA approach to describe the thermal 
emission of a finite object (both MO and non-MO). Moreover, we use our generalized approach to provide 
a rigorous demonstration of the Kirchhoff law relating the emissivity and absorptivity of a finite MO 
object. Finally, we also provide the correct expression for the radiative heat transfer between MO 
dipoles, which is the basis to study many-body effects in systems of non-reciprocal particles. Our 
work focuses on the analysis of homogeneous MO objects with constant temperatures, but it can be 
straightforwardly generalized to more complicated situations including the description of 
nonhomogeneous materials or situations with complicated temperature profiles. Our work provides a 
practical method to investigate many interesting effects in the context of the thermal emission and 
NFRHT between MO systems. 

The rest of the paper is organized as follows. In section \ref{sec-DDA} we review the DDA 
method for MO systems that we use as a starting point for our generalized TDDA approach. Section 
\ref{sec-dipoles} is devoted to the description of the radiative heat transfer between MO dipoles. 
In section \ref{sec-TDDA} we describe in detail our generalized TDDA approach for the description 
of the radiative heat transfer between two MO bodies. In section \ref{sec-TE} we show how the
TDDA can be used to describe the total thermal emission of an arbitrary finite object. In section 
\ref{sec-Kirchhoff} we address the issue of the Kirchhoff law of thermal radiation of MO objects. 
In particular, we derive here the definition of the directional, polarization-dependent emissivity 
that has to be compared with the corresponding absorption cross section to prove Kirchhoff's law. 
In section \ref{sec-results} we present the numerical results obtained for the thermal radiation 
and radiative heat transfer between different objects of arbitrary size and shape. The goal of this
section is to illustrate both the validity and the capabilities of the formalisms developed in the previous 
sections. We conclude the paper in section \ref{sec-conclusions} with some additional remarks 
about our TDDA approach and the main conclusions of our work. On the other hand, we have included 
two appendixes where we briefly discuss the convergence of the method (Appendix A) and provide an 
alternative derivation of the main result of section \ref{sec-TDDA} (Appendix B). 

\section{DDA for magneto-optical systems: A reminder} \label{sec-DDA}

Our formulation of the TDDA is based on a DDA extension to describe the electromagnetic
response of MO systems that has been recently put forward by one of us \cite{deSousa2016}.
To make this paper more self-contained, we present here a brief description of this method,
which in turn will allow us to illustrate the peculiarities of the DDA approach for MO
objects.

Let us recall that a MO system is characterized by a non-symmetric permittivity tensor
whose components depend on an external magnetic field or on the magnetization state. 
With this idea in mind, we consider an optically anisotropic finite object characterized 
by a spatially-dependent dielectric permittivity tensor $\hat \epsilon(\mathbf r)$ that is embedded 
in a homogeneous medium, which hereafter is assumed to be vacuum. In the absence of currents inside 
the object, the electric field is given by the solution of the volume-integral equation \cite{Novotny2012}
\begin{equation}
\label{eq-VIE}
\mathbf E(\mathbf r) = \mathbf E_0(\mathbf r) + k^2_0 \int_{V} \hat G(\mathbf r, \mathbf r^{\prime})
[\hat \epsilon(\mathbf r) -\hat 1] \mathbf E(\mathbf r^{\prime}) d\mathbf r^{\prime} .
\end{equation}
Here, $\mathbf E_0(\mathbf r)$ is source electric field or the field in the absence 
of the object, $k_0 = \omega/c$ is the magnitude of the vacuum wave vector, $V$ is the volume 
of the object, and $\hat G(\mathbf r, \mathbf r^{\prime})$ is the vacuum dyadic Green tensor 
given by \cite{Novotny2012}
\begin{eqnarray}
\hat G(\mathbf r, \mathbf r^{\prime}) & = & \frac{e^{ik_0R}}{4\pi R} \left[ \left( 1 +
\frac{ik_0R -1}{k^2_0 R^2} \right) \hat 1 + \right. \nonumber \\ & & \left. 
\left( \frac{3 - 3ik_0R -k^2_0 R^2}{k^2_0 R^2} \right) 
\frac{\mathbf R \otimes \mathbf R}{R^2} \right] ,
\end{eqnarray}
where $\mathbf R = \mathbf r - \mathbf r^{\prime}$, $R = |\mathbf r - \mathbf r^{\prime}|$,
and $\otimes$ denotes the exterior product.

In the DDA approach, the previous integral equation is solved by discretizing the volume $V$ as
$V= \sum^N_{n=1} V_n$, where $V_n$ is the volumen of an homogeneous region where the electric
field is assumed to be constant. Thus, Eq.~(\ref{eq-VIE}) now reads
\begin{equation}
\label{eq-VIE-DDA}
\mathbf E(\mathbf r) = \mathbf E_0(\mathbf r) + k^2_0 \sum_n \hat G(\mathbf r, \mathbf r_n)
\left[ \hat \epsilon(\mathbf r) -\hat 1 \right] \mathbf E(\mathbf r_n) V_n .
\end{equation}

Defining the dipole moments as
\begin{equation}
\mathbf p_n = \epsilon_0 V_n \left[ \hat \epsilon(\mathbf r) -\hat 1 \right] \mathbf E(\mathbf r_n) ,
\end{equation}
we can rewrite Eq.~(\ref{eq-VIE-DDA}) as
\begin{equation}
\label{eq-VIE-DDA-p}
\mathbf E(\mathbf r) = \mathbf E_0(\mathbf r) + \frac{k^2_0}{\epsilon_0} \sum_n 
\bar{\hat G}(\mathbf r, \mathbf r_n) \mathbf p_n ,
\end{equation}
where
\begin{equation}
\bar{\hat G}(\mathbf r, \mathbf r_n) = \frac{1}{V_n} \int_{V_n} \hat G(\mathbf r, \mathbf r^{\prime})
d\mathbf r^{\prime} .
\end{equation}

It can be shown that \cite{deSousa2016}
\begin{equation}
k^2_0 \bar{\hat G}(\mathbf r, \mathbf r_n) \approx
k^2_0 \hat G(\mathbf r, \mathbf r_n) \;\; \mbox{if} \; \mathbf r \notin V_n 
\end{equation}
and
\begin{eqnarray}
k^2_0 \bar{\hat G}(\mathbf r, \mathbf r_n) & \approx & 
-\hat L_n/V_n + i k^2_0 \mbox{Im} \{ \hat G(\mathbf r_n, \mathbf r_n) \} = \\
& &  -\hat L_n/V_n + i k^3_0/(6\pi) \hat 1 \;\; \mbox{if} \; \mathbf r \in V_n .
\end{eqnarray}
Here, $\hat L_n$ is the so-called electrostatic depolarization dyadic
that depends on the shape of the volume element $V_n$ \cite{Lakhtakia1992,Yaghjian1980}. 
For the case of a parallelepiped of volume $V_n = L_{n,x} L_{n,y} L_{n,z}$, $\hat L_n$ 
adopts the form \cite{Yaghjian1980}
\begin{equation}
\left[ \hat L_n \right]_{ij} = \delta_{ij} \frac{2}{\pi} 
\arctan \left( \frac{1}{L^2_{n,i}}
\frac{V_n}{\sqrt{L^2_{n,x} + L^2_{n,y} + L^2_{n,z}}} \right) . \nonumber
\end{equation}

Thus, we can now rewrite Eq.~(\ref{eq-VIE-DDA-p}) for the internal field, 
$\mathbf E_n \equiv \mathbf E(\mathbf r_n)$, as follows
\begin{eqnarray}
\label{eq-En}
\left[ \hat 1 + \left( \hat L_n - iV_n \frac{k^3_0}{6\pi} \right) [\hat \epsilon_n
- \hat 1 ] \right] \mathbf E_n & = &  \mathbf E_{0,n} + \nonumber \\
k^2_0 \sum_{m \neq n} \hat G_{nm} [\hat \epsilon(\mathbf r_m) - \hat 1] V_m 
\mathbf E_m , & &
\end{eqnarray}
where $\mathbf E_{0,n} \equiv \mathbf E_0(\mathbf r_n)$, $\hat \epsilon_n \equiv 
\hat \epsilon(\mathbf r_n)$ and $\hat G_{nm} \equiv \hat G(\mathbf r_n, \mathbf r_m)$.

The left hand side of Eq.~(\ref{eq-En}) can be defined as the exciting field
$\mathbf E_{\rm exc}(\mathbf r_n)$, i.e., the field that excites the $n$-volumen element.
Now, defining the polarizability tensor of the $n$-volumen element, $\hat \alpha_n$, as
\begin{equation}
\label{eq-alpha}
\hat \alpha_n = \left( \hat \alpha_{0,n}^{-1} - i \frac{k^3_0}{6\pi} \hat 1 \right)^{-1} ,
\end{equation}
where
\begin{equation}
\label{eq-alpha0}
\hat \alpha_{0,n}^{-1} = \frac{1}{V_n} \left( \hat L_n + [\hat \epsilon_n - \hat 1]^{-1} 
\right)
\end{equation}
is the quasistatic polarizability tensor, Eq.~(\ref{eq-En}) can be rewritten as 
a set of coupled dipole equations for the exciting fields at each element
\begin{equation}
\label{eq-Exc}
\mathbf E_{{\rm exc},n} = \mathbf E_{0,n} + 
k^2_0 \sum_{m \neq n} \hat G_{nm} \hat \alpha_m \mathbf E_{{\rm exc},m} .
\end{equation}

It is worth stressing that this DDA formulation includes automatically the so-called 
radiative corrections \cite{Sipe1974,Belov2003,Albaladejo2010}, which are related to the 
imaginary part of the Green tensor, and it is thus fully consistent with the optical 
theorem. On the other hand, from the solution of Eq.~(\ref{eq-Exc}), which constitutes 
a set of 3$N$ coupled linear equations for the exciting fields, one can get the dipole
moments and the total internal fields as follows
\begin{eqnarray}
\label{eq-pn}
\mathbf p_n & = & \epsilon_0 \hat \alpha_n \mathbf E_{{\rm exc},n} \\
\label{eq-En-pn}
\mathbf E_n & = & \frac{1}{\epsilon_0 V_n} [\hat \epsilon_n - \hat 1]^{-1} 
\mathbf p_n .
\end{eqnarray}

Let us conclude this section with a few useful remarks. First, for cubic volumen 
elements, the depolarization tensor is diagonal: $\hat L_n = (1/3) \hat 1$. For 
volumen elements of spherical or cubic, optically isotropic materials, the 
polarizability tensor is diagonal: $\hat \alpha_n = \alpha_n \hat 1$ with
\begin{equation}
\label{eq-alpha-sphere}
\alpha_n = \frac{\alpha_{0,n}}{1- ik^3_0 \alpha_{0,n}/(6\pi)},\,\,
\alpha_{0,n} = 3 V_n \left( \frac{\epsilon_n -1}{\epsilon_n + 2} \right).
\end{equation}
Finally, from the knowledge of the dipole moments and the internal fields, one
can easily compute the different cross sections (scattering, absorption, and
extinction). In particular, the absorption cross section, which will play 
an important role later on, can be obtained as follows. Assuming a plane-wave
illumination, $\mathbf E_0(\mathbf r) = \mathbf E_0 e^{i\mathbf k_0 \cdot \mathbf r}$,
the absorption cross section is given by \cite{deSousa2016}
\begin{equation}
\label{eq-Cabs1}
\sigma_{\rm abs} = \frac{k_0}{\epsilon^2_0 |\mathbf E_0|^2} \sum_n 
\mbox{Im} \left\{\mathbf p_n \cdot [\hat \alpha^{-1}_{0,n} \mathbf p_n ]^{\ast} \right\} .
\end{equation}

\section{Radiative heat transfer between magneto-optical dipoles} \label{sec-dipoles}

Before presenting our generalized formulation of the TDDA for arbitrary bodies,
it is instructive to first discuss the radiative heat transfer between MO dipoles. 
This discussion will allow us to highlight the peculiarities of MO systems.

The problem that we address in this section is the radiative heat transfer
(both in the near and in the far field) between two MO particles that are small compared
to their thermal wavelengths such that they can treated as point electrical dipoles.
Let us assume that these two dipoles (or dipolar particles) are located in positions
$\mathbf r_1$ and $\mathbf r_2$, they have arbitrary polarizability tensors $\hat \alpha_1$ 
and $\hat \alpha_2$, and they are at temperatures $T_1$ and $T_2$. To compute the net power 
exchanged between the two dipoles, we first compute the power dissipated in dipole 2 due to 
the emission from dipole 1, $P_{1 \to 2}$, assuming that dipole 2 does not emit \cite{note1}. 
This power is given by
\begin{equation}     
P_{1 \to 2} = \int_{V_2} \langle \mathbf j_2(\mathbf r,t) \cdot \mathbf E(\mathbf r, t)
\rangle d\mathbf r = \langle \frac{d\mathbf p_2(t)}{dt} \cdot 
\mathbf E_2(t) \rangle ,
\end{equation}
where $\mathbf j_2(\mathbf r,t) = \frac{d\mathbf p_2(t)}{dt}$ is the local electric current 
density in the volume $V_2$ and $\mathbf E(\mathbf r, t)$ is the local electric field at 
position $\mathbf r$. Moreover, $\langle \cdot \cdot \cdot \rangle$ denotes the statistical 
average that takes into account the stochastic nature of the dipoles. Let us recall that
the thermal emission originates from the fluctuating part of the dipole moments.

Now, we can express $\mathbf p_2(t)$ and $\mathbf E_2(t)$ in terms of their Fourier
transforms
\begin{equation}
\mathbf p_2(t) = \int^{\infty}_{-\infty} \frac{d\omega}{2\pi} \mathbf p_2(\omega)
e^{-i \omega t}, \,
\mathbf E_2(t) = \int^{\infty}_{-\infty} \frac{d\omega}{2\pi} \mathbf E_2(\omega)
e^{-i \omega t} \nonumber .
\end{equation}
Using the fact that these two functions are real, one can easily show that
\begin{eqnarray}
\label{eq-Im1}
\frac{d\mathbf p_2(t)}{dt} \cdot \mathbf E_2(t) & = & 2 \int^{\infty}_0 
\frac{d\omega}{2\pi} \omega \int^{\infty}_{-\infty} \frac{d\omega^{\prime}}{2\pi} \\
& & \times \mbox{Im} \left\{ \mathbf p_2(\omega) \cdot \mathbf E^{\ast}_2(\omega^{\prime})
e^{-i (\omega - \omega^{\prime}) t} \right\} \nonumber . 
\end{eqnarray}
Since the fluctuation-dissipation theorem (FTD) will introduce a $\delta$-function of
the type $\delta(\omega - \omega^{\prime})$, we focus on the calculation of 
$\mbox{Im} \left\{ \langle \mathbf p_2(\omega) \cdot \mathbf E^{\ast}_2(\omega) 
\rangle \right\}$ and in most cases we shall drop the argument $\omega$ to alleviate
the notation.

In order to determine the dipole moment and field, we need to solve the DDA equations,
see Eq.~(\ref{eq-Exc}). The exciting field at the position of dipole 2 is given by
\begin{equation}
\label{eq-Exc2}
\mathbf E_{\rm exc,2} = \mathbf E_{0,2} + k^2_0 \hat G_{21} \hat \alpha_1 
\mathbf E_{\rm exc,1}
\end{equation}
To close this equation, we need the corresponding equation for $\mathbf E_{\rm exc,1}$,
which reads
\begin{equation}
\label{eq-Exc1}
\mathbf E_{\rm exc,1} = k^2_0 \hat G_{12} \hat \alpha_2 \mathbf E_{\rm exc,2}.
\end{equation}
Notice that there is no source term in this case. Introducing Eq.~(\ref{eq-Exc1}) in 
Eq.~(\ref{eq-Exc2}), we arrive at
\begin{equation}
\mathbf E_{\rm exc,2} = \hat D_{22} \mathbf E_{0,2} , 
\end{equation}
where
\begin{equation}
\hat D_{22} = \left[\hat 1 - k^4_0 \hat G_{21} \hat \alpha_1
\hat G_{12} \hat \alpha_2 \right]^{-1} .
\end{equation}
The source field $\mathbf E_{0,2}$, due to the emission of the fluctuating part
of dipole 1, $\mathbf p_{{\rm f},1}$, is given by $\mathbf E_{0,2} = (k^2_0/\epsilon_0) 
\hat G_{21} \mathbf p_{{\rm f},1}$. Thus, 
\begin{equation}
\label{eq-Eexc2-p1}
\mathbf E_{\rm exc,2} = \frac{k^2_0}{\epsilon_0} \hat C_{21} \mathbf p_{{\rm f},1} ,
\end{equation}
where we have defined $\hat C_{21} = \hat D_{22} \hat G_{21}$. 

Now, using Eqs.~(\ref{eq-pn}) and (\ref{eq-En-pn}), we can express the dipole
moment $\mathbf p_2$ and the internal field $\mathbf E_2$ in terms of 
$\mathbf E_{\rm exc,2}$ as 
\begin{eqnarray}
\mathbf p_2 & = & \epsilon_0 \hat \alpha_2 \mathbf E_{{\rm exc},2} \\
\mathbf E_2 & = & \frac{1}{\epsilon_0 V_2} [\hat \epsilon_2 - \hat 1]^{-1}
\hat \alpha_2 \mathbf E_{{\rm exc},2} .
\end{eqnarray}
To compute $\mbox{Im} \left\{ \langle \mathbf p_2 \cdot \mathbf E^{\ast}_2 \rangle \right\}$ 
it is convenient to use a matrix notation where column vectors like $\mathbf p_2$
are understood as ($3 \times 1$) matrices and row vectors like $\mathbf p^T_2$ are 
($1 \times 3$) matrices. With this notation, we get rid of the dot product (or scalar 
product) in favor of matrix multiplications as follows
\begin{eqnarray}
\label{eq-mat}
\hspace*{-0.5cm} \mbox{Im} \left\{\langle \mathbf p_2 \cdot \mathbf E^{\ast}_2 \rangle \right\} = 
\mbox{Im} \left\{\langle \mathbf p^T_2 \mathbf E^{\ast}_2  \rangle \right\} =
\mbox{Im} \mbox{Tr} \left\{ \langle \mathbf E^{\dagger}_2 \mathbf p_2 \rangle \right\} & = & \nonumber \\ 
\frac{\epsilon_0}{V_2} \mbox{Im} \mbox{Tr} \left\{\langle \mathbf E_{{\rm exc},2} 
\mathbf E^{\dagger}_{{\rm exc},2} \rangle \hat \alpha^{\dagger}_2 
[\hat \epsilon^{\dagger}_2 - \hat 1]^{-1} \hat \alpha_2  \right\} . & &
\end{eqnarray}
Now, using Eqs.~(\ref{eq-alpha}) and (\ref{eq-alpha0}) it is easy to show that
\begin{equation}
\hat \alpha^{\dagger}_2 [\hat \epsilon^{\dagger}_2 - \hat 1]^{-1} \hat \alpha_2 =
V_2 \left[ \hat \alpha_2 - \hat \alpha^{\dagger}_2 \left( \hat L^{\dagger}_2/V +
ik^3_0/(6\pi) \hat 1 \right) \hat \alpha_2 \right] . \nonumber
\end{equation}
Introducing this relation into Eq.~(\ref{eq-mat}) and after several simple algebraic
manipulations, we arrive at
\begin{equation}
\mbox{Im} \left\{\langle \mathbf p_2 \cdot \mathbf E^{\ast}_2 \rangle \right\} =
\epsilon_0 \mbox{Tr} \left\{\langle \mathbf E_{{\rm exc},2}  
\mathbf E^{\dagger}_{{\rm exc},2} \rangle \hat \chi_2 \right\} ,
\end{equation}
where 
\begin{equation}
\label{eq-chi}
\hat \chi_i = \frac{1}{2i} \left( \hat \alpha_i - \hat \alpha^{\dagger}_i \right)
- \frac{k^3_0}{6\pi} \hat \alpha^{\dagger}_i \hat \alpha_i .
\end{equation}
Notice that $\hat \chi_i = \hat \chi^{\dagger}_i$.

Now, making use of Eq.~(\ref{eq-Eexc2-p1}), we obtain 
\begin{equation}
\label{eq-Im2}
\mbox{Im} \left\{\langle \mathbf p_2 \cdot \mathbf E^{\ast}_2 \rangle \right\} =
\frac{k^4_0}{\epsilon_0} \mbox{Tr} \left\{\hat C_{21} \langle \mathbf p_{{\rm f}, 1}
\mathbf p^{\dagger}_{{\rm f}, 1} \rangle \hat C^{\dagger}_{21} \hat \chi_2 \right\} .
\end{equation}

The statistical average appearing in the previous equation can be determined 
with the help of the fluctuation-dissipation theorem (FDT), which in its most general
form reads \cite{Landau1980,Keldysh1994,Joulain2005}
\begin{equation}
\label{eq-FDT-p1}
\langle \mathbf p_{{\rm f},1} (\omega) \mathbf p^{\dagger}_{{\rm f},1} (\omega^{\prime}) \rangle =
\hbar \epsilon_0 2 \pi \delta(\omega-\omega^{\prime}) [1 + 2n_{\rm B}(\omega, T_1)]
\hat \chi_1 ,
\end{equation}
where $n_{\rm B}(\omega,T) = [\exp(\hbar \omega/k_{\rm B}T) -1 ]^{-1}$ is the
Bose function. Several remarks are pertinent at this stage. First, notice that
the FDT of Eq.~(\ref{eq-FDT-p1}) involves $\hat \chi_1$, which contains two
terms, see Eq.~(\ref{eq-chi}). The first one is the standard contribution, while the 
second one is related to the radiative correction and its contribution avoids the 
violation of the optical theorem \cite{Manjavacas2012}. The origin of this term
has been nicely explained by Messina \emph{et al.}\ \cite{Messina2013} and the
result above is a generalization of their arguments to the MO case. On the other hand, 
notice that in the first term in the expression of $\hat \chi$ the combination 
$(\hat \alpha - \hat \alpha^{\dagger})/(2i)$ appears, which for MO systems differs 
from $\mbox{Im} \{\hat \alpha \}$.
This latter combination was used in the work of Ref.~[\onlinecite{Nikbakht2014}] 
for the description of radiative heat transfer between anisotropic particles and
therefore, that formulation is not valid for MO systems, as it was assumed in
the work on the photon thermal Hall effect of Ref.~[\onlinecite{Ben-Abdallah2016}].  

Now, we can combine Eqs.~(\ref{eq-FDT-p1}), (\ref{eq-Im2}), and (\ref{eq-Im1}) to
write the power dissipated in dipole 2 due to the emission from dipole 1 as follows
\begin{equation}
P_{1 \to 2} = 2 \int^{\infty}_0 \frac{d\omega}{2\pi} \hbar \omega 
[1+2n_{\rm B}(\omega,T_1)] k^4_0 \mbox{Tr} \left\{\hat C_{21} \hat \chi_1 
\hat C^{\dagger}_{21} \hat \chi_2 \right\} .
\end{equation}

Following the same line of reasoning, one can show that the power absorbed
by the particle 1 due to the emission of particle 2 is given by
\begin{equation}
P_{2 \to 1} = 2 \int^{\infty}_0 \frac{d\omega}{2\pi} \hbar \omega 
[1+2n_{\rm B}(\omega,T_2)] k^4_0 \mbox{Tr} \left\{\hat C_{12} \hat \chi_2 
\hat C^{\dagger}_{12} \hat \chi_1 \right\} ,
\end{equation}
where $\hat C_{12}$ can be obtained from $\hat C_{21}$ by interchanging the 
indexes 1 and 2. 

It is straightforward to show that
\begin{equation}
\mbox{Tr} \left\{\hat C_{21} \hat \chi_1 \hat C^{\dagger}_{21} \hat \chi_2 \right\} =
\mbox{Tr} \left\{\hat C_{12} \hat \chi_2 \hat C^{\dagger}_{12} \hat \chi_1 \right\} .
\end{equation}
Thus, the net power exchange, $P_{\rm net} = P_{1 \to 2}- P_{2 \to 1}$, between 
the two particles is given by the Landauer-like formula 
\begin{equation}
\label{eq-Pnet-dip}
P_{\rm net} = \int^{\infty}_0 \frac{d\omega}{2\pi} 
[\Theta (\omega,T_1) - \Theta (\omega,T_2)] {\cal T}(\omega) ,
\end{equation}
where $\Theta(\omega,T) = \hbar \omega n_{\rm B}(\omega,T)$ and ${\cal T}(\omega)$
is the frequency-dependent transmission function given by
\begin{equation}
\label{eq-T-dip}
{\cal T}(\omega) = 4 k^4_0 \mbox{Tr} \left\{\hat C_{21} \hat \chi_1 
\hat C^{\dagger}_{21} \hat \chi_2 \right\} .
\end{equation}

We emphasize that this result differs from that reported in 
Refs.~[\onlinecite{Nikbakht2014,Ben-Abdallah2016}] and it only coincides with those
results for non-MO particles. It is also important to remark that in the case of 
isotropic particles, it reduces to the result reported in the literature in which 
radiative corrections have been taken into account \cite{Manjavacas2012,Messina2013}.

\section{Generalized TDDA: Radiative heat transfer between two arbitrary
magneto-optical bodies} \label{sec-TDDA}

In this section we present our generalized TDDA for the description of the radiative
heat transfer between MO objects of arbitrary shape and for arbitrary separations. 
In particular, we focus here on the case of two finite objects assumed to be at fixed 
temperatures $T_1$ and $T_2$ and do not consider the interaction with any thermal bath. 

In the spirit of the DDA, we assume that these two bodies are described by a collection 
of $N_1$ (body 1) and $N_2$ (body 2) electrical point dipoles. Each dipole is characterized 
by a volumen $V_{i,b}$ and a polarizability tensor $\hat \alpha_{i,b}$, where $b=1,2$ 
indicates to which body the dipole belongs and $i=1, \dots ,N_1$ if $b=1$ and $i=1, \dots, 
N_2$ if $b=2$. The information about the individual dipole moments and the internal electrical 
field can be grouped into column supervectors (denoted with a bar) as follows
\begin{equation}
\bar{\mathbf P} = \left( \begin{array}{cc} \bar{\mathbf P}_1 \\ \bar{\mathbf P}_2 \end{array} \right); \;\;
\bar{\mathbf P}_1 = \left( \begin{array}{ccc} \mathbf p_{1,1} \\ \vdots \\ \mathbf p_{N_1,1} \end{array} \right),
\; \bar{\mathbf P}_2 = \left( \begin{array}{ccc} \mathbf p_{1,2} \\ \vdots \\ \mathbf p_{N_2,2} \end{array} \right)  
\end{equation}
\begin{equation}
\bar{\mathbf E} = \left( \begin{array}{cc} \bar{\mathbf E}_1 \\ \bar{\mathbf E}_2 \end{array} \right); \;\;
\bar{\mathbf E}_1 = \left( \begin{array}{ccc} \mathbf E_{1,1} \\ \vdots \\ \mathbf E_{N_1,1} \end{array} \right),
\; \bar{\mathbf E}_2 = \left( \begin{array}{ccc} \mathbf E_{1,2} \\ \vdots \\ \mathbf E_{N_2,2} \end{array} \right) .
\end{equation}
The same notation will be used for the exciting and source electric fields.

The calculation of the net radiative power exchanged by arbitrary objects is a straightforward
generalization of the calculation for two dipoles presented in the previous section 
and we shall follow the same strategy. Thus, we first compute the power absorbed by object 2
due to the thermal emission of object 1, $P_{1 \to 2}$, which is given by   
\begin{eqnarray}     
P_{1 \to 2} & = & \langle \frac{d\bar{\mathbf P}_2(t)}{dt} \cdot \bar{\mathbf E}_2(t) \rangle = \\
& & \hspace*{-1cm} 2 \int^{\infty}_0 \frac{d\omega}{2\pi} \omega \int^{\infty}_{-\infty} 
\frac{d\omega^{\prime}}{2\pi} \mbox{Im} \left\{ \langle \bar{\mathbf P}_2(\omega) \cdot \bar{\mathbf 
E}^{\ast}_2(\omega^{\prime}) \rangle e^{-i (\omega - \omega^{\prime}) t} \right\} \nonumber . 
\end{eqnarray}

Following the previous section, the goal is now to compute the exciting field 
$\bar{\mathbf E}_{{\rm exc},2}$. For this purpose, we use Eq.~(\ref{eq-Exc}), which can be 
rewritten in a more compact form using the notation above
\begin{equation}
\bar{\mathbf E}_{{\rm exc}} = \bar{\mathbf E}_0 + 
k^2_0 \Delta \bar G \bar \alpha \bar{\mathbf E}_{{\rm exc}} .
\end{equation}
Here, $\Delta \bar G = \bar G - \mbox{diag}(\bar G)$ with
\begin{equation}
\bar G = \left( \begin{array}{cc} \bar G_{11} & \bar G_{12} \\ 
\bar G_{21} & \bar G_{22} \end{array} \right) \; \mbox{and} \;
\bar \alpha = \left( \begin{array}{cc} \bar \alpha_1 & 0 \\ 
0 & \bar \alpha_2 \end{array} \right) ,
\end{equation}
where $[\bar{G}_{b,b^{\prime}}]_{ij}$ is the vacuum dyadic Green tensor connecting 
dipole $i$ in body $b$ with dipole $j$ in body $b^{\prime}$ and 
$\bar \alpha_b = \mbox{diag}(\hat \alpha_{1,b}, \dots, \hat \alpha_{N_b,b})$
with $b=1,2$.

Solving now for $\bar{\mathbf E}_{{\rm exc}}$, we have
\begin{equation}
\bar{\mathbf E}_{{\rm exc}} = \bar D \bar{\mathbf E}_0 ,
\end{equation}
where 
\begin{eqnarray}
\label{eq-Dbar}
\bar D & = & \left[ \bar 1 - k^2_0 \Delta \bar G \bar \alpha \right]^{-1} = \\ 
& & \left( \begin{array}{cc} \bar 1 - k^2_0 \Delta \bar G_{11} \bar \alpha_1
 & - k^2_0 \Delta \bar G_{12} \bar \alpha_2 \\ -k^2_0 \Delta \bar G_{21} 
 \bar \alpha_1 & \bar 1 - k^2_0 \Delta \bar G_{22} \bar \alpha_2 \end{array} \right)^{-1} .
 \nonumber
\end{eqnarray}

The source field, $\bar{\mathbf E}_0$, due to the emission of the fluctuating dipoles in 
body 1, $\bar{\mathbf P}_{{\rm f},1}$, is given by
\begin{equation}
\label{eq-ExcitingField}
\bar{\mathbf E}_0 = \frac{k^2_0}{\epsilon_0} \Delta \bar G \left( \begin{array}{c}
\bar{\mathbf P}_{{\rm f},1} \\ 0 \end{array} \right) = \frac{k^2_0}{\epsilon_0}
\left( \begin{array}{c} \Delta \bar G_{11} \bar{\mathbf P}_{{\rm f},1} \\ 
\Delta \bar G_{21} \bar{\mathbf P}_{{\rm f},1} \end{array} \right) .
\end{equation}
Thus,
\begin{equation}
\bar{\mathbf E}_{{\rm exc},2} = \frac{k^2_0}{\epsilon_0} \left( \bar D_{21} \Delta \bar G_{11}
+ \bar D_{22} \Delta \bar G_{21} \right) \bar{\mathbf P}_{{\rm f},1} .
\end{equation}
From Eq.~(\ref{eq-Dbar}) it is easy to show that
\begin{eqnarray}
\bar D_{22} & = & \left[ \bar 1 - k^2_0 \Delta \bar G_{22} \bar \alpha_2 - \right. \\ & &
\left. k^4_0 \Delta \bar G_{21} \bar \alpha_1 \left[ \bar 1 - k^2_0 \Delta \bar G_{11} 
\bar \alpha_1 \right]^{-1} \Delta \bar G_{12} \bar \alpha_2 \right]^{-1} , \nonumber \\
\bar D_{21} & = & k^2_0 \bar D_{22} \Delta \bar G_{21} \bar \alpha_1 \left[ \bar 1 - k^2_0 \Delta 
\bar G_{11} \bar \alpha_1 \right]^{-1} .
\end{eqnarray}
Therefore,
\begin{equation}
\bar{\mathbf E}_{{\rm exc},2} = \frac{k^2_0}{\epsilon_0} \bar C_{21} 
\bar{\mathbf P}_{{\rm f},1} ,
\end{equation}
where 
\begin{equation}
\bar C_{21} = \bar D_{22} \Delta \bar G_{21} \left[\bar 1 - k^2_0 \bar \alpha_1 
\Delta \bar G_{11} \right]^{-1} .
\end{equation}

Now, we use Eqs.~(\ref{eq-pn}) and (\ref{eq-En-pn}) to express the dipole
moment $\bar{\mathbf P}_2$ and the internal field $\bar{\mathbf E}_2$ in terms of 
$\bar{\mathbf E}_{\rm exc,2}$ as 
\begin{equation}
\bar{\mathbf P}_2 = \epsilon_0 \bar \alpha_2 \bar{\mathbf E}_{{\rm exc},2} \;\; \mbox{and} \;\;
\bar{\mathbf E}_2 = \bar \beta_2 \bar \alpha_2 \bar{\mathbf E}_{{\rm exc},2} ,
\end{equation}
where $\bar \beta_2 = \mbox{diag}( [\epsilon_0 V_{1,2}(\hat \epsilon_{1,2} - \hat 1)]^{-1},
\dots, [\epsilon_0 V_{N_2,2}(\hat \epsilon_{N_2,2} - \hat 1)]^{-1})$.

Following the same steps as in the previous section, it is straightforward to show that
\begin{eqnarray}
\langle \bar{\mathbf P}_2 \cdot \bar{\mathbf E}^{\ast}_2 \rangle & = & 
\epsilon_0 \mbox{Tr} \left\{\langle \bar{\mathbf E}_{{\rm exc},2}  
\bar{\mathbf E}^{\dagger}_{{\rm exc},2} \rangle \bar \chi_2 \right\} = \nonumber \\ & &
\frac{k^4_0}{\epsilon_0} \mbox{Tr} \left\{\bar C_{21} \langle \bar{\mathbf P}_{{\rm f}, 1}
\bar{\mathbf P}^{\dagger}_{{\rm f}, 1} \rangle \bar C^{\dagger}_{21} \bar \chi_2 \right\} ,
\end{eqnarray}
where $\bar \chi_b = \mbox{diag}(\hat \chi_{1,b}, \dots, \hat \chi_{N_b,b})$ with $b=1,2$.
Let us recall that $\hat \chi_{i,b}$ is given by Eq.~(\ref{eq-chi}). Again, the
statistical average appearing in the previous equation can be computed with the FDT,
which now reads
\begin{equation}
\label{eq-FDT-P1}
\langle \bar{\mathbf P}_{{\rm f},1} (\omega) \bar{\mathbf P}^{\dagger}_{{\rm f},1} (\omega^{\prime}) \rangle =
\hbar \epsilon_0 2 \pi \delta(\omega-\omega^{\prime}) [1 + 2n_{\rm B}(\omega, T_1)] \bar \chi_1 .
\end{equation}

Now, we can combine the last two equations to write the power dissipated in body 2 due 
to the emission from body 1 as follows
\begin{equation}
P_{1 \to 2} = 2 \int^{\infty}_0 \frac{d\omega}{2\pi} \hbar \omega 
[1+2n_{\rm B}(\omega,T_1)] k^4_0 \mbox{Tr} \left\{\bar C_{21} \bar \chi_1 
\bar C^{\dagger}_{21} \bar \chi_2 \right\} .
\end{equation}

Analogously, one can show that the power absorbed by body 1 due to the emission of 
body 2 is given by
\begin{equation}
P_{2 \to 1} = 2 \int^{\infty}_0 \frac{d\omega}{2\pi} \hbar \omega 
[1+2n_{\rm B}(\omega,T_2)] k^4_0 \mbox{Tr} \left\{\bar C_{12} \bar \chi_2 
\bar C^{\dagger}_{12} \bar \chi_1 \right\} ,
\end{equation}
where $\bar C_{12}$ can be obtained from $\bar C_{21}$ by interchanging the 
indexes 1 and 2. 

On the other hand, it can be shown that
\begin{equation}
\mbox{Tr} \left\{\bar C_{21} \bar \chi_1 \bar C^{\dagger}_{21} \bar \chi_2 \right\} =
\mbox{Tr} \left\{\bar C_{12} \bar \chi_2 \bar C^{\dagger}_{12} \bar \chi_1 \right\} .
\end{equation}
Thus, the net power exchange, $P_{\rm net} = P_{1 \to 2}- P_{2 \to 1}$, between 
the two bodies is given by the Landauer-like formula 
\begin{equation}
\label{eq-Pnet-bodies}
P_{\rm net} = \int^{\infty}_0 \frac{d\omega}{2\pi} 
[\Theta (\omega,T_1) - \Theta (\omega,T_2)] {\cal T}(\omega) ,
\end{equation}
where let us recall that $\Theta(\omega,T) = \hbar \omega n_{\rm B}(\omega,T)$ 
and ${\cal T}(\omega)$ is the transmission function given by
\begin{equation}
\label{eq-T-bodies}
{\cal T}(\omega) = 4 k^4_0 \mbox{Tr} \left\{\bar C_{21} \bar \chi_1 
\bar C^{\dagger}_{21} \bar \chi_2 \right\} .
\end{equation}
Eqs.~(\ref{eq-Pnet-bodies}) and (\ref{eq-T-bodies}) are our central result for the 
description of the radiative heat transfer between anisotropic objects of arbitrary 
shape, which is valid in particular for MO systems. Obviously, the result summarized 
in these equations reduces to that of Eqs.~(\ref{eq-Pnet-dip}) and (\ref{eq-T-dip}) 
for the case of two dipoles.

The result of Eqs.~(\ref{eq-Pnet-bodies}) and (\ref{eq-T-bodies}) is valid for 
arbitrary separations between the two bodies, i.e., it includes both the near-field
and the far-field contributions. In the far-field, the formula can be simplified 
by neglecting the multiple scattering between the two bodies. This can be 
done by approximating the matrix $\bar C_{21}$ by
\begin{equation}
\label{eq-FFapprox}
\bar C_{21} \approx \left[ \bar 1 - k^2_0 \Delta G_{22} \bar \alpha_2 \right]^{-1}
\Delta \bar G_{21} \left[\bar 1 - k^2_0 \bar \alpha_1 \Delta \bar G_{11} \right]^{-1} .
\end{equation}

Let us conclude this section by saying that our derivation of the result for the
radiative heat transfer between two finite objects is based on an intuitive division
of the problem into two subproblems in which one object is assumed to be the emitter
and the other one the receiver, and in each subproblem the receiver is assumed to
not radiate. This is indeed the strategy followed by Polder and van Hove in their
seminal paper in which they computed the radiative heat transfer between two parallel
plates \cite{Polder1971}. Anyway, one might wonder if this \emph{ad hoc} division is 
fully justified in the problem addressed in this section. To demonstrate that this is
indeed the case, we present in Appendix \ref{sec-derivation} an alternative derivation 
of the central result of this section starting from a different point of view in which 
both objects radiate ``simultaneously" and the net exchanged power is directly calculated.
This alternative derivation detailed in Appendix \ref{sec-derivation} confirms the validity 
of the central result of this section summarized in Eqs.~(\ref{eq-Pnet-bodies}) and 
(\ref{eq-T-bodies}).

\section{Thermal emission of finite object} \label{sec-TE}

The thermal emission of a finite object is another important issue that can be 
described with the help TDDA, but it has not been done so far. The goal of this 
section, and the next one, is to fill this gap. We divide our discussion of the 
formulation of thermal radiation within TDDA into two parts. The first one is 
addressed in this section, where we present a convenient way to compute the total radiation
power emitted by a finite object. The analysis of the angular and polarization
dependence of the thermal emission will be carried out in the next section, where, 
in particular, we generalize Kirchhoff's law of thermal radiation to the case of 
MO objects of arbitrary size and shape.

The problem we want to address in this section is the calculation of the 
total radiation power emitted by a finite body at temperature $T$. In this case, 
we model this body as a collection of $N$ point dipoles that interact with the
electric field of a thermal bath at temperature $T$. The total power emitted by 
the body, $P_{\rm em}$, must be equal to the total power absorbed from the bath 
when they are at the same temperature. Thus, we can follow the previous section
and write the emitted power as   
\begin{eqnarray}     
P_{\rm em} & = & \langle \frac{d\bar{\mathbf P}(t)}{dt} \cdot \bar{\mathbf E}(t) \rangle = \\
& & \hspace*{-1cm} 2 \int^{\infty}_0 \frac{d\omega}{2\pi} \omega \int^{\infty}_{-\infty} 
\frac{d\omega^{\prime}}{2\pi} \mbox{Im} \left\{ \langle \bar{\mathbf P}(\omega) \cdot \bar{\mathbf 
E}^{\ast}(\omega^{\prime}) \rangle e^{-i (\omega - \omega^{\prime}) t} \right\} \nonumber ,
\end{eqnarray}
where we are using the same type of notation as in the previous section, i.e.,
\begin{equation}
\bar{\mathbf P} = \left( \begin{array}{ccc} \mathbf p_{1} \\ \vdots \\ \mathbf p_{N} \end{array} \right), \;\;
\bar{\mathbf E} = \left( \begin{array}{ccc} \mathbf E_{1} \\ \vdots \\ \mathbf E_{N} \end{array} \right) .
\end{equation}

In this problem, the source field is the field of the bath, i.e., $\bar{\mathbf E}_0 = 
\bar{\mathbf E}_{\rm B}$, where we have used the notation of previous equation. Thus, 
the exciting field $\bar{\mathbf E}_{\rm exc}$ is simply given by 
\begin{equation}
\bar{\mathbf E}_{{\rm exc}} = \bar D \bar{\mathbf E}_{\rm B} ,
\end{equation}
where $\bar D = \left[ \bar 1 - k^2_0 \Delta \bar G \bar \alpha \right]^{-1}$. Here, 
$\Delta \bar G = \bar G - \mbox{diag}(\bar G)$, where $[\bar{G}]_{ij}$ is the vacuum dyadic 
Green tensor connecting dipole $i$ with dipole $j$ inside the body and
$\bar \alpha = \mbox{diag}(\hat \alpha_{1}, \dots, \hat \alpha_{N})$.

As in the previous section, it is straightforward to show that
\begin{eqnarray}
\langle \bar{\mathbf P} \cdot \bar{\mathbf E}^{\ast} \rangle & = & 
\epsilon_0 \mbox{Tr} \left\{\langle \bar{\mathbf E}_{\rm exc}  
\bar{\mathbf E}^{\dagger}_{\rm exc} \rangle \bar \chi \right\} = \nonumber \\ & &
\epsilon_0 \mbox{Tr} \left\{\bar D \langle \bar{\mathbf E}_{\rm B}
\bar{\mathbf E}^{\dagger}_{\rm B} \rangle \bar D^{\dagger} \bar \chi \right\} ,
\end{eqnarray}
where $\bar \chi = \mbox{diag}(\hat \chi_{1}, \dots, \hat \chi_{N})$. Now,
to compute the statistical average appearing in the previous equation, we make use
of the FDT theorem for the field correlations of the bath, which reads \cite{Messina2013}
\begin{equation}
\label{eq-FDT-Eb}
\langle \bar{\mathbf E}_{\rm B} (\omega) \bar{\mathbf E}^{\dagger}_{\rm B} (\omega^{\prime}) \rangle =
\frac{\hbar k^2_0}{\epsilon_0} 2 \pi \delta(\omega-\omega^{\prime}) [1 + 2n_{\rm B}(\omega, T)] 
\mbox{Im} \{ \bar G \} .
\end{equation}

Now, we can combine the last two equations to write the emitted power as follows
\cite{note2}
\begin{equation}
P_{\rm em} = 8\pi^2 \int^{\infty}_0 d\omega \, I_{\rm BB}(\omega,T)
\mbox{Tr} \left\{\bar D \mbox{Im} \{\bar G \}  \bar D^{\dagger} \bar \chi \right\} ,
\end{equation}
where
\begin{equation}
I_{\rm BB}(\omega,T) = \frac{\omega^2}{4\pi^3 c^2} \frac{\hbar \omega}{e^{\hbar \omega/
k_{\rm B}T} -1} 
\end{equation}
is Planck distribution function. In the previous equation one can identify 
\begin{equation}
\label{eq-emissivity}
{\cal E}(\omega) = 2\pi \mbox{Tr} \left\{\bar D \mbox{Im} \{\bar G \}  
\bar D^{\dagger} \bar \chi \right\}
\end{equation}
as a quantity that, when divided by the geometrical cross section of the body, plays the role 
of an effective angular-averaged frequency-dependent emissivity. In the case of a single
dipole, ${\cal E}(\omega) = (1/3) k_0 \mbox{Tr} \{ \hat \chi \}$, which for a 
spherical, optically isotropic dipole ($\hat \chi = \chi \hat 1$) reduces to 
${\cal E}(\omega) = k_0 \chi = k_0 [ \mbox{Im} \{ \alpha \} - k^3_0 |\alpha|^2/(6\pi) ]$,
where $\alpha$ is given by Eq.~(\ref{eq-alpha-sphere}). As we show below, the quantity
$k_0 \chi$ is simply the absorption cross section of this isotropic dipole.

\section{Kirchhoff law of thermal radiation of magneto-optical objects} \label{sec-Kirchhoff}

Another fundamental question that we want to address in this work is the validity of Kirchhoff's 
law for finite MO objects. This thermal radiation law states that the emissivity is equal to
the absorptivity. This law is a textbook result for the case of extended systems, but its 
proof for finite objects has also been reported for both isothermal bodies \cite{Rytov1989} and 
non-isothermal ones \cite{Greffet2016}. However, these proofs are restricted to reciprocal 
objects, i.e., objects with symmetric permittivity tensors, and our goal is to extend the 
analysis of the validity of this law to the case of MO (non-reciprocal) objects.

In its most general form, Kirchhoff's law involves the thermal emission in a given direction 
and for a given polarization. Thus, our first task is to find a proper expression for the
polarization-dependent directional emissivity. For this purpose, it is convenient to write 
the total emitted power in a way different than above, i.e., in terms of Poynting's vector as
\begin{equation}
P_{\rm em} =  \int_A \langle \mathbf S(\mathbf r, t) \rangle \cdot \hat{\mathbf r} \, dA,
\end{equation}
which describes the integrated flux across a differential section $dA=R^2 \sin\theta d\theta d\phi$ 
perpendicular to a radial unit vector $\hat{\mathbf r} = \mathbf R/R$, performed over a sphere 
of radius $R$ enclosing the emitting object. Since this quantity does not depend on the actual 
choice of $R$, for convenience, we will evaluate Poynting's vector in the far field.

Now, we need to compute the statistical average of Poynting's vector resulting from the thermal 
emission of an arbitrary body, i.e., $\langle \mathbf S(\mathbf r, t) \rangle = \langle 
\mathbf E(\mathbf r, t) \times \mathbf H(\mathbf r, t) \rangle$, where $\mathbf r$ is 
the point of observation (outside the object) and $\mathbf E(\mathbf r, t)$ and 
$\mathbf H(\mathbf r, t)$ are the electric and magnetic field, respectively. As usual,
we can express this average in terms of the Fourier transforms of the fields as
\begin{eqnarray}     
\langle \mathbf S(\mathbf r, t) \rangle & = &  \\ && \hspace*{-1.6cm}
2 \int^{\infty}_0 \frac{d\omega}{2\pi} \int^{\infty}_{-\infty} 
\frac{d\omega^{\prime}}{2\pi} \mbox{Re} \left\{ \langle \mathbf E(\mathbf r, \omega) \times \mathbf 
H^{\ast}(\mathbf r, \omega^{\prime}) \rangle e^{-i (\omega - \omega^{\prime}) t} \right\} \nonumber.
\end{eqnarray}
Anticipating that the FDT theorem will introduce a $\delta$-function of the type 
$\delta(\omega - \omega^{\prime})$, we shall focus on the calculation of $\mbox{Re} 
\{ \langle \mathbf E(\mathbf r, \omega) \times \mathbf H^{\ast}(\mathbf r, \omega) \rangle \}$
and we shall not write explicitly the argument $\omega$.

It is worth noting that $\langle \mathbf S(\mathbf r, t) \rangle\cdot\hat{\mathbf r}$, in 
virtue of the superposition principle, can always be described in terms of two linearly 
polarized fields that are perpendicular to each other, both lying on a plane perpendicular 
to $\hat{\mathbf r}$, as $\langle \mathbf S(\mathbf r, t) \rangle\cdot\hat{\mathbf r} = 
\langle S_r\rangle_{e1}+\langle S_r\rangle_{e2}$.

In the spirit of the TDDA, we assume that our body is described by a collection of $N$ point
dipoles located at positions $\mathbf r_n$. Using the notation of previous sections, the electric 
field generated by these dipoles is given by 
\begin{equation}
\mathbf E(\mathbf r) = \mathbf E_0(\mathbf r) + \frac{k^2_0}{\epsilon_0} \bar{\mathbf G} \bar{\mathbf P} ,
\end{equation}
where we have defined the following row supervector of Green tensors
\begin{equation}
\label{eq-SuperGvector}
\bar{\mathbf G} = \left(\hat G(\mathbf r, \mathbf r_1), \dots, \hat G(\mathbf r, \mathbf r_N) 
\right) .
\end{equation}
As usual, $\mathbf E_0(\mathbf r)$ is the source field that originates from the fluctuating
dipoles inside that body and is given by 
\begin{equation}
\mathbf E_0(\mathbf r) = \frac{k^2_0}{\epsilon_0} \bar{\mathbf G} \bar{\mathbf P}_{\rm f} .
\end{equation}
The total dipole moments $\bar{\mathbf P}$ can be obtained from the exciting field inside
the body as $\bar{\mathbf P} =\epsilon_0 \bar \alpha \bar{\mathbf E}_{\rm exc}$, while 
$\bar{\mathbf E}_{\rm exc}$ fulfills the DDA equation
\begin{equation}
\bar{\mathbf E}_{\rm exc} = \bar{\mathbf E}_0 + k^2_0 \Delta \bar G \bar \alpha \bar{\mathbf E}_{\rm exc},
\end{equation}
where $\bar{\mathbf E}_0 = (k^2_0/\epsilon_0) \Delta \bar G \bar{\mathbf P}_{\rm f}$.
Solving for $\bar{\mathbf E}_{\rm exc}$, we obtain that
\begin{equation}
\bar{\mathbf E}_{\rm exc} = \frac{k^2_0}{\epsilon_0} (\bar 1 - k^2_0 \Delta \bar G \bar \alpha)^{-1}
\Delta \bar G \bar{\mathbf P}_{\rm f}
\end{equation}
from which it is straightforward to show that the total field is given by
\begin{equation}
\label{eq-Etotal}
\mathbf E(\mathbf r) = \frac{k^2_0}{\epsilon_0} \bar{\mathbf G} \bar T^{-1} \bar{\mathbf P}_{\rm f},
\end{equation}
where
\begin{equation}
\bar T = \bar 1 - k^2_0 \bar \alpha \Delta \bar G .
\end{equation}

In the far-field region, the magnetic field is simply related to the electric field via
\begin{equation}
\mathbf H (\mathbf r) = Z^{-1}_0 \hat{\mathbf r} \times \mathbf E(\mathbf r),
\end{equation}
where $Z^{-1}_0 = \epsilon_0 c$% and $\hat{\mathbf r} = \mathbf r/r$
. 

Now, to select
the contribution of a given polarization, we write the electric field as $\mathbf 
E(\mathbf r) = E_e \hat{\mathbf e}$, where $\hat{\mathbf e}$ is a unit vector defining
the polarization state and lies in a plane perpendicular to the radial direction $\hat{\mathbf r}$. In
our matrix notation, the field amplitude reads $E_e = \hat{\mathbf e}^T \mathbf E(\mathbf r)$
and we have
\begin{equation}
\langle \mathbf E(\mathbf r) \times \mathbf{H}^{\ast}(\mathbf r) \rangle = 
Z^{-1}_0 \langle E_e E^{\ast}_e \rangle \hat{\mathbf r} ,
\end{equation}
which with the help of Eq.~(\ref{eq-Etotal}) can be written as
\begin{eqnarray}
\langle \mathbf E(\mathbf r) \times \mathbf{H}^{\ast}(\mathbf r) \rangle & = & \\
& & \hspace*{-1cm} \frac{k^4_0 c}{\epsilon_0} \mbox{Tr} \left\{\bar{\mathbf G} \bar T^{-1} 
\langle \bar{\mathbf P}_{\rm f} \bar{\mathbf P}^{\dagger}_{\rm f} \rangle 
\bar T^{-1 \dagger} \bar{\mathbf G}^{\dagger} \hat{\mathbf e} 
\hat{\mathbf e}^T \right\} \hat{\mathbf r} . \nonumber
\end{eqnarray}

Now, we can make use of the FDT to write
\begin{equation}
\langle \bar{\mathbf P}_{\rm f} (\omega) \bar{\mathbf P}^{\dagger}_{\rm f} (\omega^{\prime}) \rangle =
\hbar \epsilon_0 2 \pi \delta(\omega-\omega^{\prime}) [1 + 2n_{\rm B}(\omega, T)] \bar \chi ,
\end{equation}
which leads to the following expression for the radial component of Poynting's vector 
for a polarization given by the unit vector $\hat{\mathbf e}$
\begin{equation} 
\label{eq-Sr}   
\langle S_r \rangle_e = 8\pi^2 \int^{\infty}_0 d\omega I_{\rm BB}(\omega,T) k_0 
\mbox{Tr} \left\{ \bar{\mathbf G} \bar T^{-1} \bar \chi \bar T^{-1 \dagger} 
\bar{\mathbf G}^{\dagger} \hat{\mathbf e} \hat{\mathbf e}^T \right\} .
\end{equation}
Let us stress that in the previous expression one must use the far-field expression of the
Green tensors, i.e., $\hat G(\mathbf r, {\mathbf r}_j)= \exp(ik_0R_j) (1 - \hat{\mathbf r}_j
\otimes \hat{\mathbf r}_j)/(4\pi R_j)$ where ${\mathbf R}_j = \mathbf r - {\mathbf r}_j$,
$R_j=|\mathbf r - {\mathbf r}_j|$, and $\hat{\mathbf r}_j = \mathbf R_j/R_j$.

Taking into account the standard definition of the total emitted power
\begin{eqnarray}
\label{eq-TotPowerEmitted-Emissiv}
P_{\rm em} & = & 4\pi \int^{\infty}_0 d\omega I_{\rm BB}(\omega,T){\cal E}(\omega) \nonumber \\
& = & \int_A dA \left(\langle S_r\rangle_{e1}+\langle S_r\rangle_{e2}\right) ,
\end{eqnarray}
we can write
\begin{eqnarray}
\label{eq-TotEmPowerAngPol}
& &P_{\rm em}  = 4\pi \int^{\infty}_0 d\omega I_{\rm BB}(\omega,T)\int_A dA \frac{1}{2} \times
\nonumber \\ & &\left\{4\pi\sum_{i=1,2} \left(k_0 \mbox{Tr} \left\{ \bar{\mathbf G} \bar T^{-1} 
\bar \chi \bar T^{-1 \dagger} \bar{\mathbf G}^{\dagger} \hat{\mathbf e_i} \hat{\mathbf e_i}^T 
\right\}\right) \right\} .
\end{eqnarray}

Equation~(\ref{eq-TotEmPowerAngPol}) can be written as
\begin{equation}   
P_{\rm em}  = 4\pi\int^{\infty}_0 d\omega I_{\rm BB}(\omega,T)\frac{1}{8\pi r^2}
\int_A dA \sum^2_{i=1}{\cal E}_{e_i}(\hat{\mathbf r},\omega)
 ,
 \label{eq-TotEmissiondA}
\end{equation}
where ${\cal E}_{e_i}(\hat{\mathbf r},\omega)$ is the polarization-dependent directional emissivity
given by 
\begin{equation}   
\label{eq-zeta-emissivity}
{\cal E}_{e_i}(\hat{\mathbf r},\omega) = (4 \pi r)^2 k_0 \mbox{Tr} \left\{\bar{\mathbf G} \bar T^{-1} 
\bar \chi \bar T^{-1 \dagger} \bar{\mathbf G}^{\dagger} \hat{\mathbf e_i} \hat{\mathbf e_i}^T \right\}.
\end{equation}
Let us stress that, as defined here, this emissivity has dimensions of area like the absorption cross
section. On the other hand, it is worth noting that Eq.~(\ref{eq-TotEmissiondA}) permits establishing the 
following relationship between the total emissivity ${\cal E}(\omega)$ and the polarization-dependent 
directional emissivity ${\cal E}_{e_i}(\hat{\mathbf r},\omega)$
\begin{equation}   
{\cal E}(\omega)  = \frac{1}{8\pi r^2}\int_A dA \sum_{i=1,2}{\cal E}_{e_i}(\hat{\mathbf r},\omega)
 ,
 \label{eq-E-Eei}
\end{equation}
where it is easy to verify that in the case of isotropic $ \chi $ it reduces to ${\cal E}(\omega) =
{\cal E}_{e_i}(\hat{\mathbf r},\omega)$.

The quantity ${\cal E}_{e_i}(\hat{\mathbf r},\omega)$ has to be compared with the absorption cross 
section for the same polarization and incident angle to verify Kirchhoff's law. The expression for the
absorption cross section is given by Eq.~(\ref{eq-Cabs1}), which can be written as
\begin{equation}
\label{eq-Cabs2}
\sigma_{\rm abs}(\omega) = \frac{k_0}{|\mathbf E_0|^2} \mbox{Tr} \left\{\bar D 
\bar{\mathbf E}_0 \bar{\mathbf E}^{\dagger}_0 \bar D^{\dagger} \bar \chi \right\} ,
\end{equation}
where
\begin{equation}
\bar{\mathbf E}_0 = \mathbf E_0 \left( \begin{array}{ccc} e^{i \mathbf k_0 \cdot \mathbf r_1}
 \\ \vdots \\ e^{i \mathbf k_0 \cdot \mathbf r_N} \end{array} \right) 
\end{equation}
is a linearly polarized plane wave $\mathbf E_0 = E_0 \hat{\mathbf e_i}$ impinging as $\hat{\mathbf k}_0 = -\hat{\mathbf r}$.

A general demonstration of Kirchhoff's law from Eqs.~(\ref{eq-zeta-emissivity}) and (\ref{eq-Cabs2}) 
is cumbersome. In what follows we will provide evidence of this fulfillment in
the case of single dipoles whereas a numerical verification for finite objects will be provided in section 
\ref{sec-results}.

In the case of a single dipole we have that $\bar T^{-1} = \bar 1$ in 
Eq.~(\ref{eq-zeta-emissivity}) and $\bar D = \hat 1$ in Eq.~(\ref{eq-Cabs2}). Then, using 
$\mathbf E_0 = E_0 \hat{\mathbf e}$, the absorption cross section can be written as $\sigma_{\rm abs}(\omega) = 
k_0 \mbox{Tr} \left\{\hat{\mathbf e} \hat{\mathbf e}^T \hat \chi \right\}$. On the other hand, if we assume
that the dipole is located at the origin of the spherical coordinate system, the Green tensor in far-field 
approximation is given by $\hat G = \exp(ik_0 R) (1 - \hat{\mathbf r} \otimes \hat{\mathbf r})/(4\pi R)$ and 
Eq.~(\ref{eq-zeta-emissivity}) becomes
\begin{eqnarray}
{\cal E}_e(\hat{\mathbf r},\omega) &&= k_0 \mbox{Tr} \left\{(1 - \hat{\mathbf r} \otimes \hat{\mathbf r}) 
\hat \chi (1 - \hat{\mathbf r} \otimes \hat{\mathbf r}) \hat{\mathbf e} \hat{\mathbf e}^T \right\}\nonumber \\ 
&&=k_0 \mbox{Tr} \left\{\hat{\mathbf e} \hat{\mathbf e}^T \hat \chi \right\}= \sigma_{\rm abs}(\omega).
\label{eq-Kirchhoff1Dip}
\end{eqnarray}

Finally, for an isotropic system this leads to the well-known ${\cal E}(\omega) = 
\sigma_{\rm abs}(\omega)$ Kirchhoff's result.

\section{Numerical results} \label{sec-results}

In this section we present a series of results related to the different aspects of radiative 
heat transfer and thermal emission discussed in previous sections. Our goal is to illustrate
the capabilities of the TDDA method put forward in this work, as well as to demonstrate its
validity by comparing with established results in the literature. For these purposes, we
shall consider below two materials. First, as an example of a non-MO material we choose 
SiO$_2$, which is polar dielectric that has been amply studied in the context of NFRHT.
We use the dielectric function of SiO$_2$ reported in Palik's book \cite{Palik1985}.
As an example of a MO material we shall consider $n$-doped InSb, a polar semiconductor, that
when subjected to an external magnetic field becomes MO. For the sake of concreteness, we shall 
assume that magnetic field is applied in the $z$-direction, $\mathbf H = H_z \mathbf z$. In this case, 
the permittivity tensor of InSb adopts the following form \cite{Palik1976}
\begin{equation}
\label{perm-tensor-Hz}
\hat \epsilon(H) = \left( \begin{array}{ccc} \epsilon_1(H) & -i\epsilon_2(H) & 0 \\
i \epsilon_2(H) & \epsilon_1(H) & 0 \\ 0 & 0 & \epsilon_3 \end{array} \right) ,
\end{equation}
where
\begin{eqnarray}
\epsilon_1(H) & = & \epsilon_{\infty} \left( 1 + \frac{\omega^2_L - \omega^2_T}{\omega^2_T - 
\omega^2 - i \Gamma \omega} + \frac{\omega^2_p (\omega + i \gamma)}{\omega [\omega^2_c -
(\omega + i \gamma)^2]} \right) , \nonumber \\
\label{eq-epsilons}
\epsilon_2(H) & = & \frac{\epsilon_{\infty} \omega^2_p \omega_c}{\omega [(\omega + i \gamma)^2 -
\omega^2_c]} , \\
\epsilon_3 & = & \epsilon_{\infty} \left( 1 + \frac{\omega^2_L - \omega^2_T}{\omega^2_T -
\omega^2 - i \Gamma \omega} - \frac{\omega^2_p}{\omega (\omega + i \gamma)} \right) . \nonumber
\end{eqnarray}
Here, $\epsilon_{\infty}$ is the high-frequency dielectric constant, $\omega_L$ is the longitudinal
optical phonon frequency, $\omega_T$ is the transverse optical phonon frequency, $\omega^2_p =
ne^2/(m^{\ast} \epsilon_0 \epsilon_{\infty})$ defines the plasma frequency of free carriers of density 
$n$ and effective mass $m^{\ast}$, $\Gamma$ is the phonon damping constant, and $\gamma$ is the 
free-carrier damping constant. Finally, the magnetic field enters in these expressions via the cyclotron 
frequency $\omega_c = eH/m^{\ast}$. Let us point out that in this case the magneto-optics is induced
by the magnetic field, which modifies the diagonal elements of the permittivity tensor and introduces
off-diagonal terms. There are two contributions to the diagonal components, namely, optical phonons
and free carriers. These latter ones are responsible for the magneto-optics at finite magnetic field.
Notice that we have neglected the contribution from intra-band transitions because we are interested
in thermal properties at room temperature, where this contribution does not play any role. 
In all calculations below, we consider the particular case taken from Ref.~[\onlinecite{Palik1976}],
where $\epsilon_{\infty} = 15.7$, $\omega_L = 3.62 \times 10^{13}$ rad/s, $\omega_T = 3.39\times 10^{13}$ 
rad/s, $\Gamma = 5.65 \times 10^{11}$ rad/s, $\gamma = 3.39 \times 10^{12}$ rad/s, $n = 1.07 \times 10^{17}$ 
cm$^{-3}$, $m^{\ast}/m = 0.022$, and $\omega_p = 3.14 \times 10^{13}$ rad/s. As a reference, let us
say that with these parameters $\omega_c = 8.02 \times 10^{12}$ rad/s for a field of 1 T.

The explicit form for the dielectric tensor given by Eq.~(\ref{perm-tensor-Hz}) gives rise to a 
$\hat\chi$, see Eq.~(\ref{eq-chi}), having the form
\begin{equation}
\label{chi-tensor-Hz}
\hat \chi = \left( \begin{array}{ccc} \chi_{xx} & \chi_{xy} & 0 \\
-\chi_{xy} & \chi_{xx} & 0 \\ 0 & 0 & \chi_{zz} \end{array} \right) .
\end{equation}

For simplicity, the results presented below are for homogeneous bodies, i.e., with a spatially
constant permittivity tensor, and for constant temperatures inside the bodies. Moreover, all the 
results for finite systems were obtained by discretizing the objects in terms of cubes of equal size. 
The corresponding polarizability tensors of the cubes were computed with Eqs.~(\ref{eq-alpha}) and 
(\ref{eq-alpha0}) with $\hat L_n = (1/3) \hat 1$, which in the case of isotropic materials
reduce to Eq.~(\ref{eq-alpha-sphere}). Unless stated otherwise, the calculations of the radiative
thermal conductance, power emission, etc., were carried out at $T=300$ K.

\subsection{Radiative heat transfer} \label{sec-results-RHT}

\begin{figure}[t]
\begin{center} 
\includegraphics[width=\columnwidth,clip]{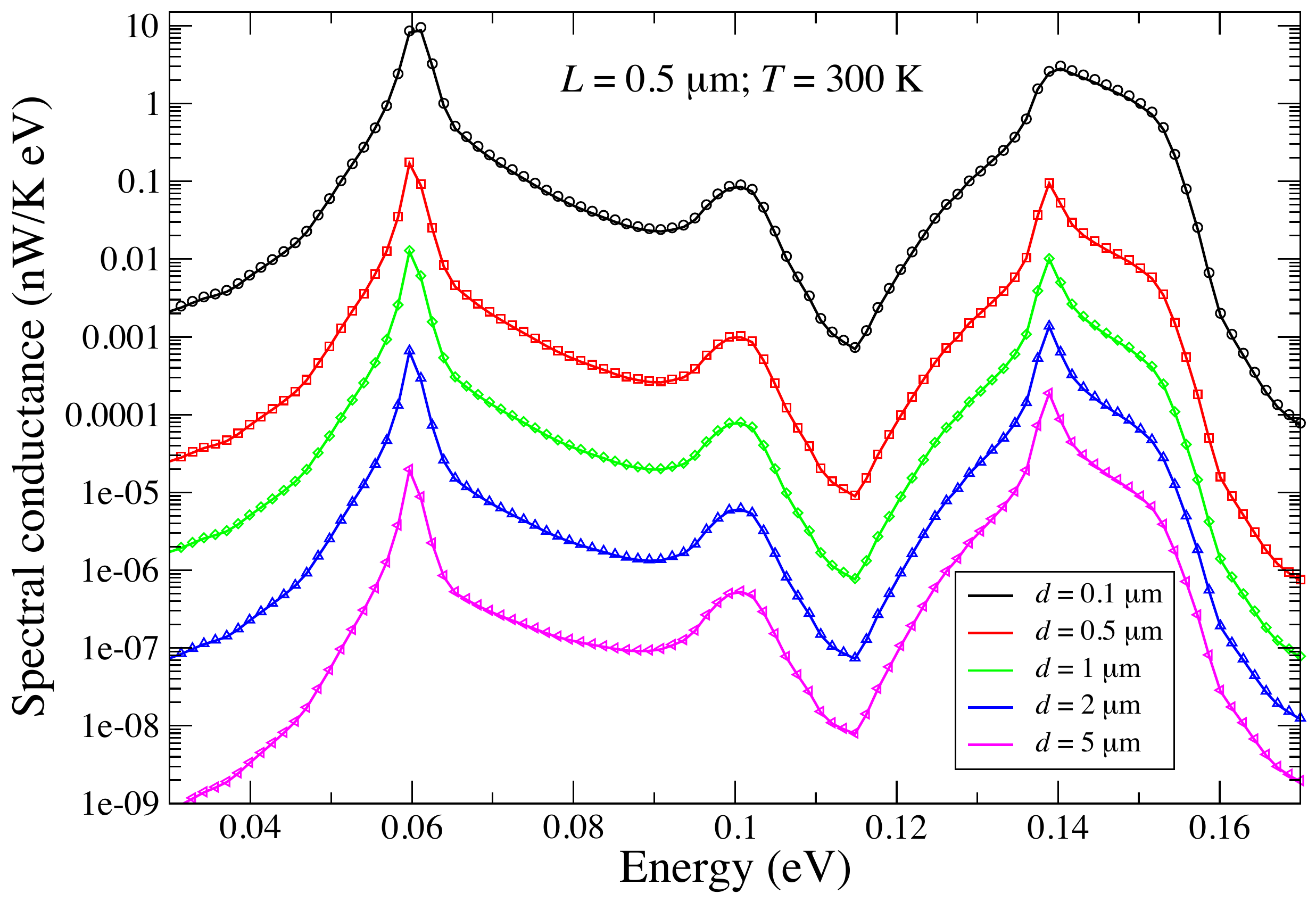} 
\end{center}
\caption{(Color online) Spectral conductance as a function of the energy for two SiO$_2$ cubes of
side $L=0.5$ $\mu$m and various gaps, $d$, at $T=300$ K. The solid lines correspond to the results computed
with the TDDA approach with 6859 dipoles per cube, while the open symbols correspond to the results 
obtained with SCUFF-EM.}
\label{fig-NFRHT-comparison1}
\end{figure}

The issue that we want to address now is the description of the radiative heat transfer between two finite
objects using the formalism detailed in section \ref{sec-TDDA}. To test the validity of this formalism,
we first consider the case of optically isotropic objects, a case that can be described with existent methods.
Indeed, in this case our formalism basically reduces to the TDDA put forward in Ref.~[\onlinecite{Edalatpour2015}], 
which was thoroughly tested against the solution for two spheres \cite{Narayanaswamy2008}. 
In our case, and in order to avoid known problems related to the so-called shape error, an error due to 
the description of objects that cannot be exactly represented by cubical lattices \cite{Draine1988,Yurkin2006}, 
we consider here the heat transfer between two identical cubes of side $L$. In Fig.~\ref{fig-NFRHT-comparison1} 
we show as solid lines the TDDA results for the spectral radiative thermal conductance (conductance per unit of 
energy) as a function of the photon energy for two SiO$_2$ cubes of side $L=0.5$ $\mu$m and various separations 
ranging from 100 nm to 5 $\mu$m. To check the validity of our TDDA approach, we compare our results with 
those obtained with the code SCUFF-EM \cite{Reid2015,SCUFF-EM}, which are shown in Fig.~\ref{fig-NFRHT-comparison1} 
as open symbols. The SCUFF-EM solver implements the fluctuating-surface-current (FSC) formulation of the heat 
transfer problem put forward in Refs.~[\onlinecite{Rodriguez2012,Rodriguez2013}] in combination with the boundary 
element method (BEM). The FSC-BEM combination used in SCUFF-EM enables the description of the radiative heat 
transfer between homogeneous, optically isotropic bodies of arbitrary shape and provide numerically exact 
results within the framework of fluctuational electrodynamics in the local approximation, i.e., assuming that 
the dielectric function only depends on frequency.   

As it can be seen in Fig.~\ref{fig-NFRHT-comparison1}, our TDDA method is able to reproduce the exact results 
obtained with SCUFF-EM. Let us say that the TDDA results shown here were obtained modeling each cube with 6859 
cubic dipoles (we briefly discuss the convergence of these results with the number of dipoles used in the 
simulations in Appendix~\ref{sec-comparison}). It is worth stressing that it 
becomes progressively more demanding to converge the TDDA results upon reducing the gap between the cubes. Thus 
for instance, in this example the difference between the TDDA and the SCUFF-EM results for the total conductance 
(integrated over energy) is of 2.5\% for a gap of $d=5$ $\mu$m, while it monotonically increases up to 6.2\% for 
$d=100$ nm. The physical reason for this behavior is the fact that the NFRHT in this case is dominated by surface 
phonon polaritons \cite{Mulet2002} that have a penetration depth comparable to the gap size \cite{Song2015a}. 
This implies that the electric field inside the cubes varies on a length scale smaller than the gap and therefore, 
one needs an increasing number of dipoles to properly describe the radiative heat transfer as the gap diminishes.
The important thing is that the error in the TDDA calculations can be systematically reduced by refining the
dipole grid, which can be done by increasing the number of dipoles or using adaptive meshes. Let us stress
that in this work we are not interested in presenting a detailed analysis of the convergence of our TDDA 
approach (since it would be strictly equivalent to that in Ref.~\cite{Edalatpour2015}), but rather in establishing 
its fundamental validity. 

\begin{figure}[t]
\begin{center} 
\includegraphics[width=\columnwidth,clip]{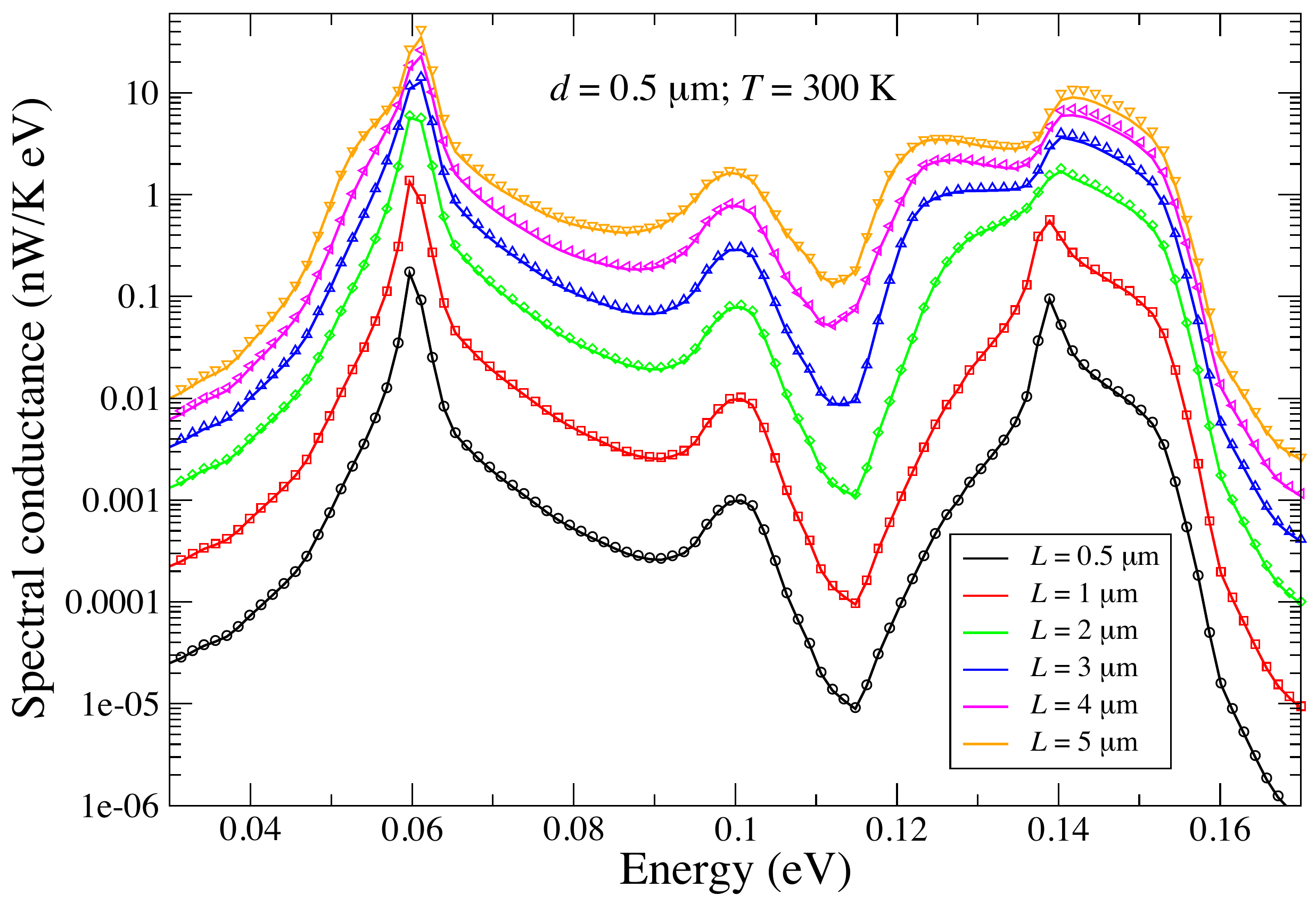} 
\end{center}
\caption{(Color online) Spectral conductance as a function of the energy for two SiO$_2$ cubes of 
different side values and separated by a gap of 0.5 $\mu$m at $T=300$ K. The solid lines correspond to 
the results computed with the TDDA approach with 6859 dipoles per cube, while the open symbols correspond 
to the results obtained with SCUFF-EM. The error in the TDDA calculations in the total conductance is
of 3.1\% for $L=0.5$ $\mu$m and it increases monotonically up to 11.3\% for $L=5$ $\mu$m.}
\label{fig-NFRHT-comparison2}
\end{figure}

We present another example of the comparison between our TDDA results and those obtained with SCUFF-EM
for the spectral conductance of two SiO$_2$ cubes of varying size and gap $d=0.5$ $\mu$m in 
Fig.~\ref{fig-NFRHT-comparison2}. Again, we find a very good agreement between both types of results,
which becomes progressively worse (for a fixed number of dipoles) as the side of the cube increases, as 
expected. More importantly, we always find that this agreement can be systematically improved by increasing 
the number of dipoles in our cubic lattices (for more details on the convergence of these results, see 
Appendix~\ref{sec-comparison}). So in short, we conclude that our TDDA method produces the correct results 
for the radiative heat transfer provided that a sufficiently large number of dipoles is employed in the 
discretization of the objects. 

\begin{figure}[t]
\begin{center} \includegraphics[width=\columnwidth,clip]{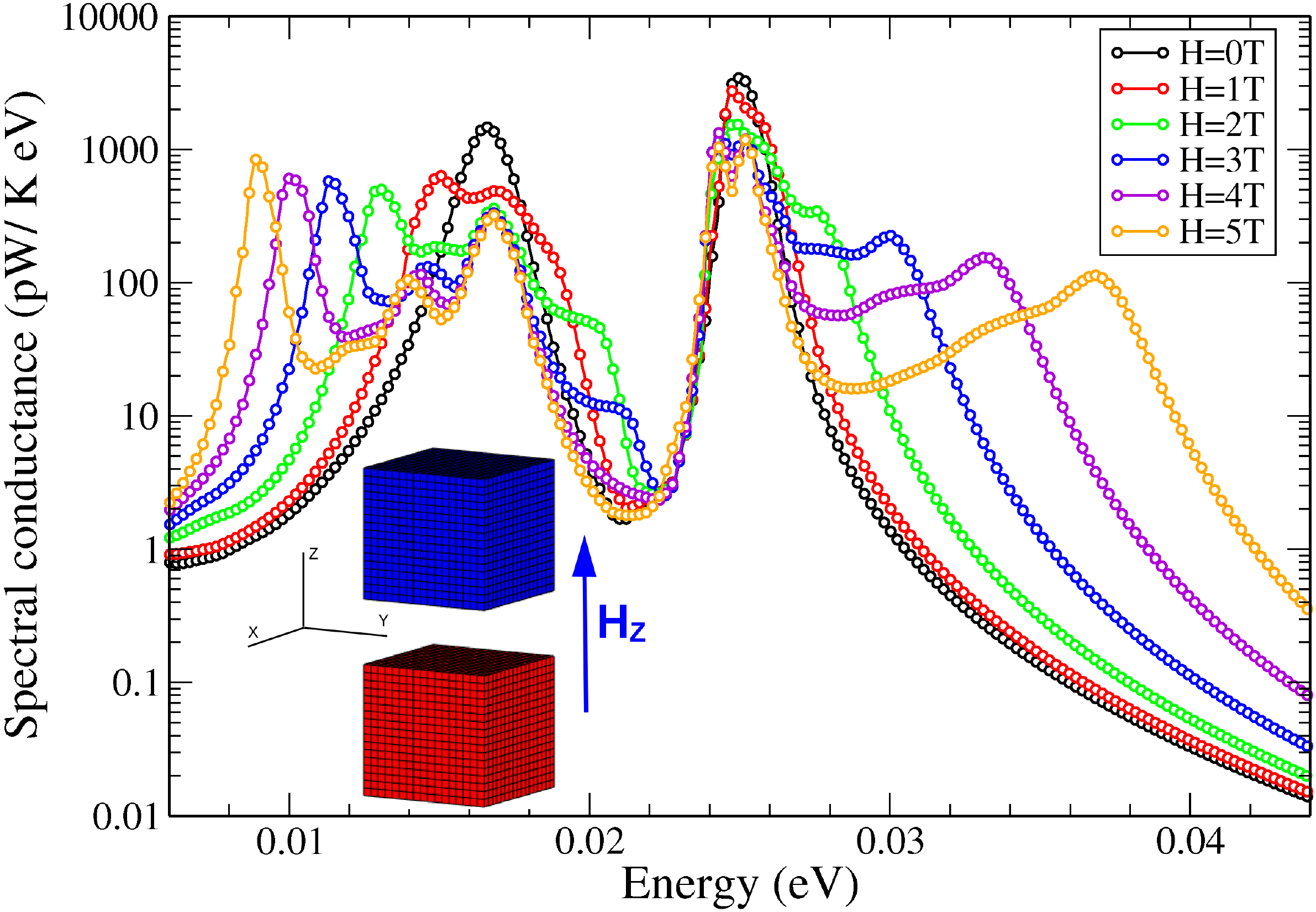} 
\end{center}
\caption{(Color online) Spectral conductance as a function of the energy for InSb cubes with 
a cube side of $1$ $\mu$m separated by a gap of 500 nm, at $T=300$ K, and for various values of 
the magnetic field, $H$, applied along the $z$-direction. The inset shows the discretization geometry, 
where the number of dipoles per cube is 4913 (each one has an edge of 59 nm).}
\label{fig-RHT_2CubesInSb}
\end{figure}

We now proceed to illustrate our approach in the case of MO objects, which is out of the scope of 
existent methods. For this purpose, we consider
the radiative heat transfer between two identical InSb cubic particles of side $L=1$ $\mu$m separated 
$d=500$ nm using 4913 cubic dipoles for each particle. The results for the spectral conductance 
for different values of the external magnetic field, $H$, are shown in Fig.~\ref{fig-RHT_2CubesInSb}.
In this case, the magnetic field is oriented as shown in the inset of this figure. As one can see,
the spectral conductance in the absence of field ($H=0$) is dominated by the contribution of two 
peaks that, as explained in Ref.~\cite{Moncada-Villa2015}, are due to the contribution of surface
plasmon polaritons (low-energy peak) and surface phonon polaritons (high-energy peak). Notice that
the magnetic field induces a splitting of the low-energy peak, while it introduces a new peak at
high energies that appears at the cyclotron frequency and therefore, its position is blue-shifted
linearly with the external field. This behavior is very similar to what it was found in 
Ref.~\cite{Moncada-Villa2015} for the case of two semi-infinite InSb plates. These results will
be discussed in more detail in a forthcoming publication and we show them here just to illustrate
the capabilities of our TDDA method. 

\begin{figure}[t]
\begin{center} \includegraphics[width=\columnwidth,clip]{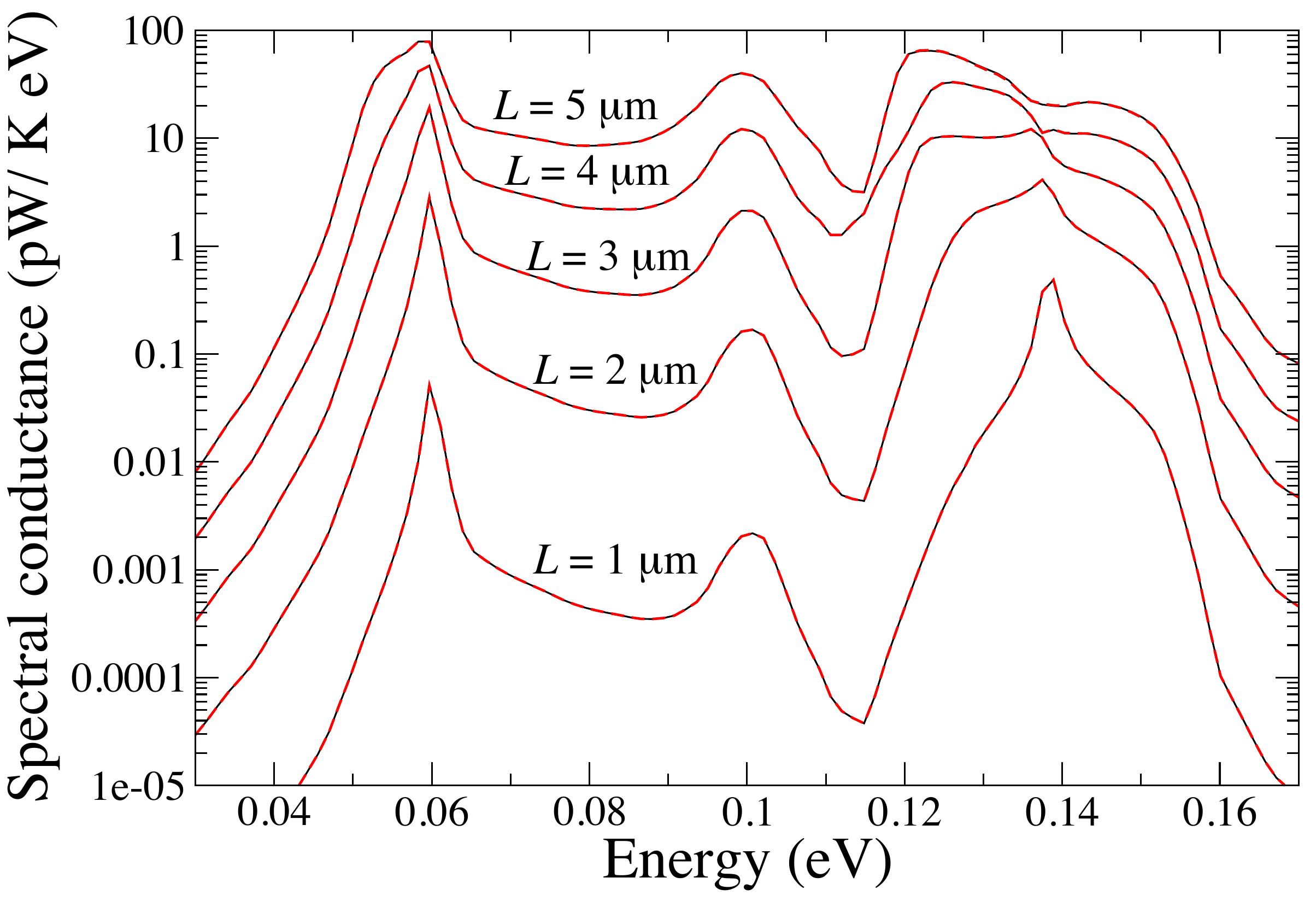} \end{center}
\caption{(Color online) Spectral conductance as a function of the energy for SiO$_2$ cubes with 
a gap of 20 $\mu$m, $T=300$ K, and for various values of the cube side, $L$. The solid lines 
correspond to the exact results and the dashed lines to the results computed with the far-field 
approximation of Eq.~(\ref{eq-FFapprox}).}
\label{fig-FF-SiO2}
\end{figure}

To conclude this subsection, we want to illustrate the validity of the far-field approximation of
Eq.~(\ref{eq-FFapprox}). For this purpose, we show in Fig.~\ref{fig-FF-SiO2} a comparison between 
the exact results and the far-field approximation for the room temperature spectral conductance for 
two identical cubes of SiO$_2$ of various sizes separated a distance of 20 $\mu$m, which is larger 
than the thermal wavelength. The results were computed with 2197 dipoles in each cube, which was
sufficient to converge the results with an accuracy of better than 1\%. As it is clear from 
Fig.~\ref{fig-FF-SiO2}, the approximation of Eq.~(\ref{eq-FFapprox}) accurately reproduces the
exact results.

\subsection{Total thermal emission} \label{sec-results-em}

\begin{figure}[t]
\begin{center} \includegraphics[width=\columnwidth,clip]{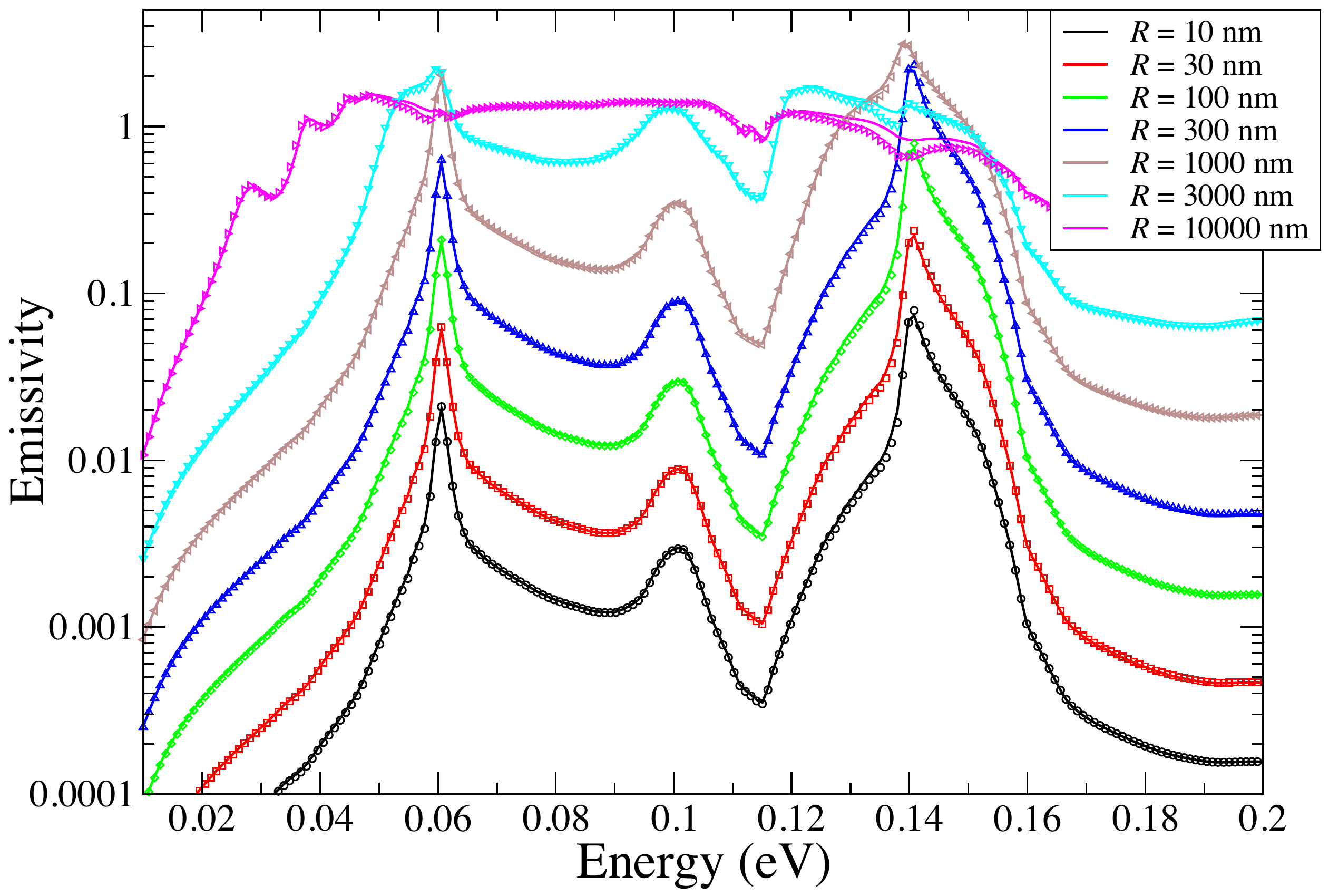} \end{center}
\caption{(Color online) Total emissivity of a SiO$_2$ sphere as a function of the energy for different
values of its radius $R$ at $T=300$ K. The solid lines correspond to the results computed with 
TDDA with $\sim 12000$ dipoles, while the open symbols correspond to the results obtained 
with Mie theory.}
\label{fig-emissivity-SiO2}
\end{figure}

In this subsection we want to discuss the total thermal emission of a finite object. To test the
validity of the formalism described in section \ref{sec-TE} we first consider the total thermal emission 
of a sphere of a non-MO material. A well-known result, first derived by Kattawar and Eisner \cite{Kattawar1970},
is that the emissivity of a non-MO sphere of arbitrary radius obtained within fluctuational electrodynamics
is equal to the corresponding absorption efficiency, i.e., to the absorption cross section normalized by 
the geometrical cross section. Let us recall that we demonstrated this result in section \ref{sec-TE} for 
the case of a single spherical dipole. Since the absorption efficiency can be computed exactly with the 
help of Mie theory \cite{Bohren1998}, the result of Kattawar and Eisner provides a very stringent test
for our theory of thermal emission. We have computed both the emissivity and absorption efficiency 
using Eqs.~(\ref{eq-emissivity}) and (\ref{eq-Cabs2}), respectively, for spheres of different sizes
and materials and in all cases we have found that they are indeed identical. In Fig.~\ref{fig-emissivity-SiO2} 
we show the TDDA results for the total emissivity of a SiO$_2$ sphere for different radii (solid lines) as 
a function of the energy. Let us clarify that we are showing the quantity defined in Eq.~(\ref{eq-emissivity})
and normalized by the geometrical cross section to make it dimensionless.
We also show the corresponding results for the absorption efficiency calculated with the exact 
Mie theory \cite{Bohren1998}. As one can see, there is a very good agreement between our TDDA results 
and the exact results, which demonstrates the validity of our formalism. Again, it is worth remarking
that as one increases the size of the sphere, a larger number of dipoles is required to satisfactorily 
converge the results. To obtain the results of Fig.~\ref{fig-emissivity-SiO2} we used about 12000
dipoles, which is enough to reproduce the exact results for spheres of radius up to approximately 10 $\mu$m.
In any case, it is important to emphasize there is no fundamental limitation to obtain the exact result
within our TDDA approach, although the practical problem to converge the results may become in some cases 
very difficult to overcome.   

\begin{figure}[t]
\begin{center} \includegraphics[width=0.95\columnwidth,clip]{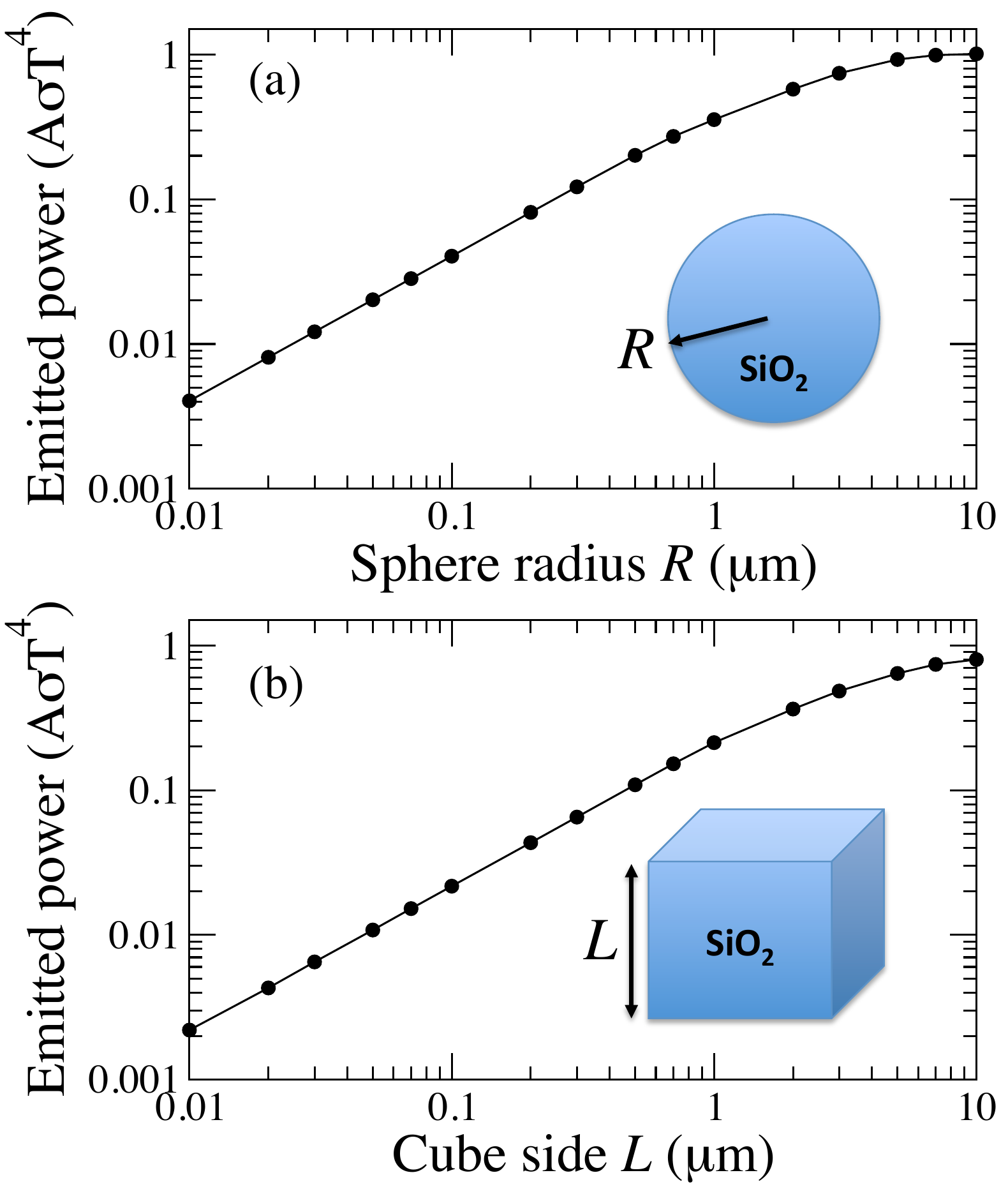} \end{center}
\caption{(Color online) (a) Total power emitted by a SiO$_2$ sphere as a function of its radius $R$
at $T=300$ K. The power is normalized with the black body result, $A \sigma T^4$, where $A$ is the
total area of the sphere. (b) The same as in panel (a), but for a SiO$_2$ cube of side $L$.}
\label{fig-total-emission-SiO2}
\end{figure}

For completeness, we show in Fig.~\ref{fig-total-emission-SiO2} the total power emitted at room temperature
by a sphere and a cube of SiO$_2$ of varying size. The results are normalized by the corresponding results
for a black body (Stefan-Boltzmann law). As one can see, for small spheres and cubes the emitted power
is proportional to the volumen of the finite objects. This is a well-known result \cite{Kruger2011,Kruger2012}, 
which is due to the fact that in this regime the skin depth of this material at the relevant wavelengths is 
larger than the characteristic size of the object. This means in practice that the whole object contributes 
to the thermal emission. However, as the size increases, the emitted power becomes proportional the area of 
the object, which reflects the fact that the size becomes larger than the skin depth and then only the surface 
of the object significantly contributes to the thermal emission. Notice also that in both cases (spheres and cubes), 
when the characteristic size becomes on the order of a few microns, the emitted power becomes comparable to 
that of a black body of the same size. Let us say that the thermal emission of a SiO$_2$ sphere was studied 
by Kr\"uger \emph{et al.}\ \cite{Kruger2011,Kruger2012} and our results agree with those reported in those 
references.

\begin{figure}[b] 
\includegraphics[width=0.95\columnwidth,clip]{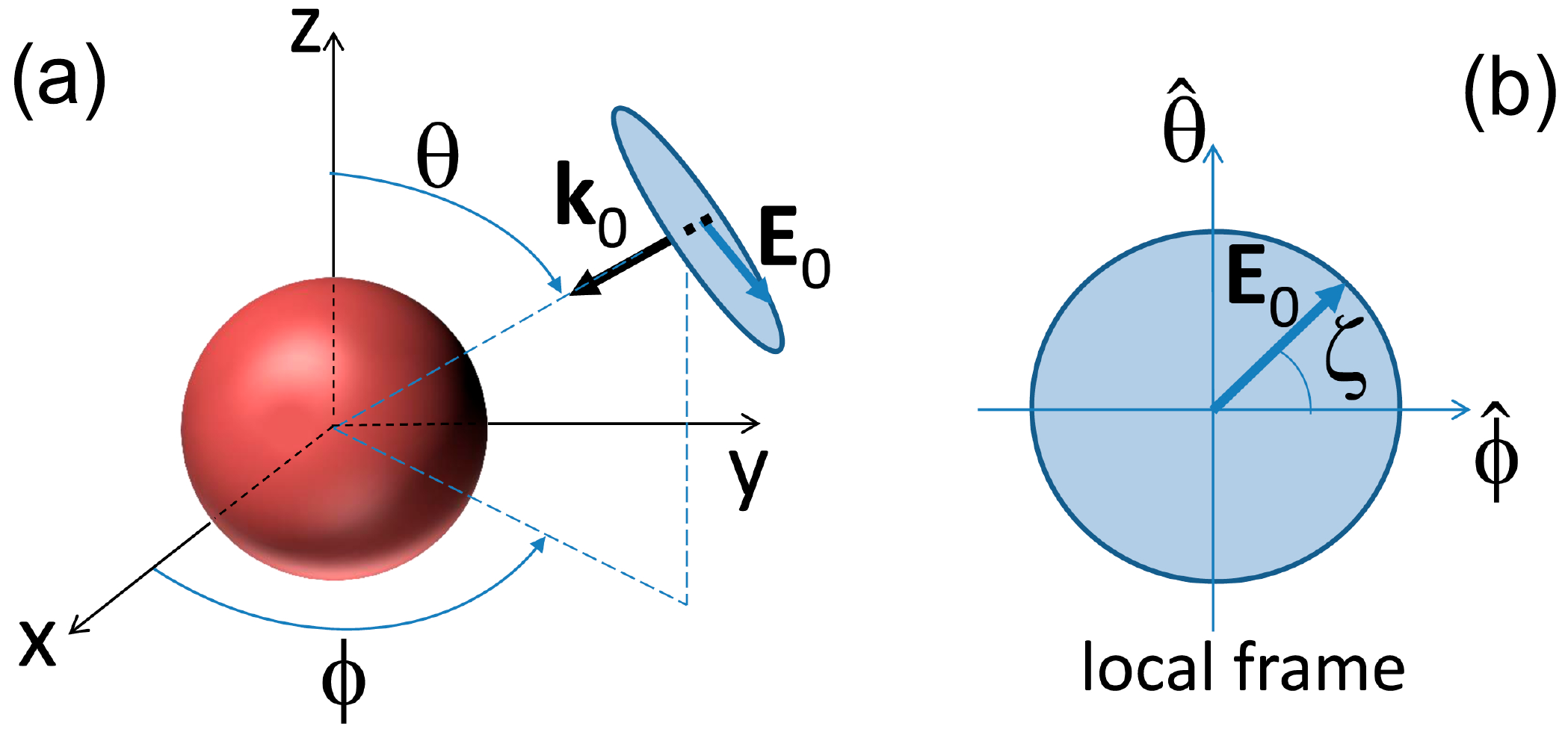}
\caption{(Color online) Schematics of the choice of coordinate system to describe the thermal emission. 
(a) Emitter in the spherical coordinate system and the plane of polarization. (b) The local frame defined 
in the plane of polarization and polarization angle.}
\label{fig-coordinatesys}
\end{figure}
\begin{figure*}[t]
\begin{center} \includegraphics[width=\textwidth,clip]{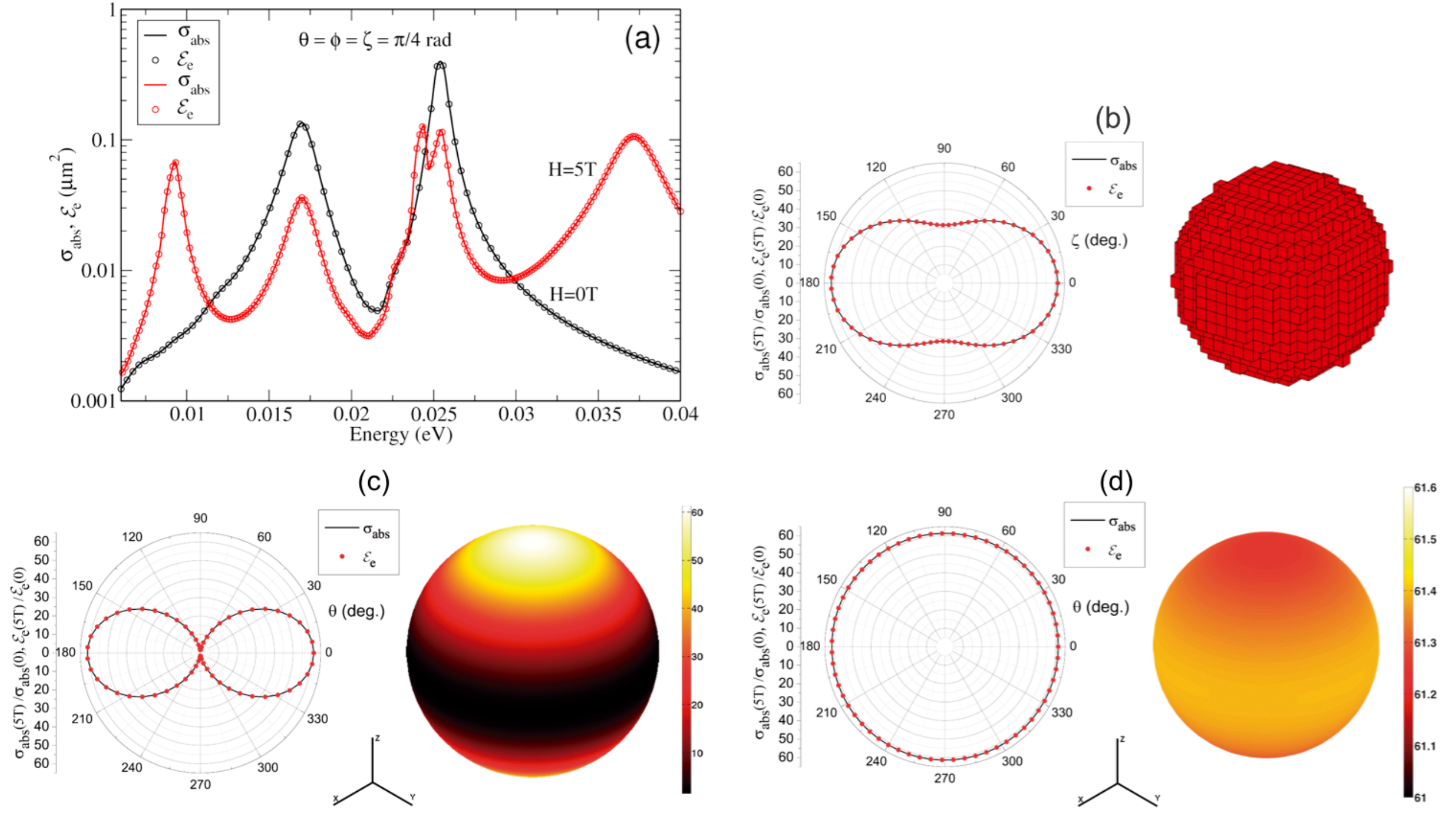} \end{center}
\caption{(Color online) Example of validity of Kirchhoff's law for an InSb sphere of radius $R=0.5$ $\mu$m 
at $T=300$ K under the action of magnetic field of $H=5$ T. (a) Comparison of the spectra of 
${\cal E}_e(\hat{\mathbf r},\omega)$ and $\sigma_{\rm abs}$ for  $\theta=\phi=\zeta=\pi/4$ rad at $H=0$ T and
$H=5$ T. (b) Dependence of ${\cal E}_e(\hat{\mathbf r},\omega)$ and $\sigma_{abs}$ on the angle of 
polarization $\zeta$ for $\hbar \omega=0.037$ eV for $H=5$ T. The inset depicts the discretization of
the sphere employed for these calculations. (c) Angular distribution and polar plot of 
${\cal E}_e(\hat{\mathbf r},\omega)$ and $\sigma_{\rm abs}$ for the $s$-polarizarion ($\zeta = \pi/2$ rad)
for $\hbar \omega=0.037$ eV and $H=5$ T, normalized to the result for $H=0$ T. (d) The same as in
panel (c) but for the $p$-polarization ($\zeta = 0$ rad). Notice that in panel (c) the independence 
on $\phi$ and the $\sin^2\theta$-dependence are clearly demonstrated, while in (d) the angular 
independence for $p$-polarization is quite apparent.}
\label{fig-kirchhofflaw}
\end{figure*}

\subsection{Kirchhoff's law for MO systems} \label{sec-results-Klaw}

The goal of this subsection is to illustrate the validity of Kirchhoff's law in the case of finite MO 
objects. But before presenting the full numerical analysis, it is instructive to consider a MO dipole 
(where we still consider MO activity is due to a magnetic field pointing along the $z$-direction) since 
it is possible to provide an analytical solution for an arbitrary polarization vector $\hat{\mathbf e}$.
Using the coordinate system depicted in Fig.~\ref{fig-coordinatesys} the spectral, polarization-dependent 
directional emissivity for a single dipole with $\hat \chi$ given by Eq.~(\ref{chi-tensor-Hz}) can be 
written for any polarization angle $\zeta$ as [see Eq.~(\ref{eq-Kirchhoff1Dip})]
\begin{eqnarray}
\label{eq-EmissivMOdipoleSimple}
&&{\cal E}_{e}(\omega,\zeta) =\sigma_{\rm abs}(\omega,\zeta) \nonumber \\
&= &k_0 \left[\chi_{xx} - (\chi_{xx} - \chi_{zz}) \sin^2\theta\sin^2\zeta\right].
\end{eqnarray}
Notice that there is no dependence on the azimuthal angle $\phi$ due to the symmetry introduced by the 
fact that $\chi_{yy}=\chi_{xx}$. Moreover, since we have to choose two orthogonal vectors, i.e., $\zeta$ and 
$\zeta+\pi/2$, the sum required to obtain the total emitted power is equal to $k_0 \left[2\chi_{xx} - 
(\chi_{xx} - \chi_{zz}) \sin^2\theta\right]$, that, as expected, does not depend on the choice of $\zeta$.

Since the direction of the external magnetic field imposes a preferential direction, it is convenient to choose 
two values of $\zeta$ so that one polarization vector has no component along the direction of the external 
magnetic field, $\zeta=0$ rad (let us call it $p$), while the other is collinear with the magnetic field, 
$\zeta=\pi/2$ rad (let us call it $s$). It is straightforward to see that ${\cal E}_{p}(\omega) =k_0\chi_{xx}$, 
i.e., independent of any incidence angle $\theta$ or $\phi$, while ${\cal E}_{s}(\omega) =k_0 \left[\chi_{xx} 
- (\chi_{xx} - \chi_{zz}) \sin^2\theta\right]$ presenting thus a $\sin^2\theta$ dependence, varying from 
${\cal E}_{s}(\omega) =k_0\chi_{xx}$ to ${\cal E}_{s}(\omega) =k_0\chi_{zz}$.  

To illustrate the fulfillment of Kirchhoff's law for a body of finite size, we consider an homogeneous InSb 
sphere of radius $R=0.5$ $\mu$m, using discretization unit cubes with an edge of $R/7=71$ nm ($\sim$ 2550 
discretization elements). In Fig.~\ref{fig-kirchhofflaw}(a) we present spectra of the polarized emissivity 
and of the absorption cross section for a direction defined by $\theta=\phi=\pi/4$ rad and polarization angle 
by $\zeta=\pi/4$ rad. Two values of magnetic field, $H=0$T and $H=5$T, are shown. For both cases 
${\cal E}_{\frac{\pi}{4}}(\frac{\pi}{4},\frac{\pi}{4},\omega) $ and $\sigma_{\rm abs}$ (for the same angles 
and polarization) are identical. 

In order to analyze the same system as a function of the geometrical angles, we verify the $\sin^2\zeta$ 
dependence obtained for a single dipole in Eq.~(\ref{eq-EmissivMOdipoleSimple}). This is clearly visible 
in Fig.~\ref{fig-kirchhofflaw}(b), where we plot ${\cal E}_{e}$ and $\sigma_{abs}$ for $\theta=\frac{\pi}{4}$ rad,
$\phi=\frac{\pi}{4}$ rad and $\omega=0.037$ eV, as a function of $\zeta$ for an applied field of $H=5$ T normalized 
to its value at $H=0$ T (let us point out that, since the geometry is fully isotropic, for $H=0$T it happens that 
$\chi_{xx}=\chi_{zz}$ and thus emissivity and absorption cross section are polarization insensitive).  

The next step in the verification is to compare the angular profile of the emissivity and of the absorption 
cross section for fixed polarizations. As it can be understood from Eq.~(\ref{eq-EmissivMOdipoleSimple}), 
the most interesting choice is the $s$-polarization ($\zeta=\pi/2$ rad). In Fig.~\ref{fig-kirchhofflaw}(c), we 
present a colormap of ${\cal E}_s(\omega) $ and $\sigma_{abs}$ (for  the same polarization) for $H=5$ T 
normalized to $H=0$ T, as a function of $\phi$ and $\theta$ showing that they are independent of $\phi$ and 
that the dependence on $\theta$ is of the form of $\sin^2\theta$, as evidenced by the associated polar plot. 
The same plot, but for $p$-polarization is presented in Fig.~\ref{fig-kirchhofflaw}(d) demonstrating that 
for this configuration and polarization, the emissivity and absorption cross section is isotropic, within 
a numerical error smaller than 2\%.

Let us conclude this subsection by saying that we have analyzed the polarization-dependent, directional
emissivity and the corresponding absorption cross section for different sizes and shapes of InSb particles 
under a magnetic field finding that in all cases these two quantities are identical. 

\section{Additional remarks and conclusions} \label{sec-conclusions}

In this work we have focused on the description of the thermal radiation and radiative heat transfer
of homogeneous objects with constant temperatures. However, all the basic formulas derived in this
work are directly applicable to the case of inhomogeneous objects, with spatially dependent dielectric
functions, and they can be straightforwardly generalized to consider the case of arbitrary temperature
profiles in the interior of the objects. On the other hand, the TDDA method presented here can easily
be generalized in a number of additional ways to treat, for instance, the radiative heat transfer 
between surfaces and finite objects \cite{Edalatpour2016} or between periodic systems \cite{Draine2008}. 

With respect to the improvement of the efficiency and accuracy of the TDDA method presented here,
there are a number of obvious improvements that one can implement without the need to modify
the formulation detailed in this work. Thus for instance, one can employ adaptive dipole lattices
to better describe the non-uniform field profiles inside the objects \cite{Edalatpour2015}, one
can discretize the objects in terms of non-cubic dipoles with shapes better adapted to the geometries
of the objects under study \cite{Smunev2015}, or one can use a DDA formulation based on the integration
of the Green's tensor (IGT) in the spirit of Ref.~\cite{Chaumet2004}, rather than on the standard 
point-dipole interaction, as done in this work.

In summary, we have presented in this work a new formulation of the TDDA approach to describe the
radiative heat between finite objects of arbitrary size and shape. This formulation allows us to 
describe the radiative heat transfer 
between MO objects and, more generally, between optically anisotropic objects with arbitrary
permittivity tensors. We have shown how this TDDA approach can also be used to describe the 
thermal emission of a finite object. Moreover, we have provided very compact and transparent 
formulas for different physical quantities that can be directly employed in generalizations of
the method presented here. In particular, we have corrected the existent formulas for the radiative
heat transfer between non-reciprocal dipoles, which can be used to analyze different many-body 
effects in ensembles of MO particles. On the other hand, we have used our TDDA approach to 
demonstrate Kirchhoff's law relating the emission and absorption of non-reciprocal objects. Our
work opens the way for the rigorous description of different thermal radiation phenomena involving
finite MO objects of arbitrary shape.    

\begin{acknowledgments}
We thank V. Fern\'andez-Hurtado for providing the numerical results obtained SCUFF-EM shown in 
this paper, and J.J. S\'aenz and N. de Sousa for stimulating discussions.
This work was financially supported by the Comunidad de Madrid (Contract No.\ S2013/MIT-2740).
J.C.C.\ also acknowledges financial support from the Spanish MINECO (Contract No.\ FIS2014-53488-P) 
and thanks the DFG and SFB 767 for sponsoring his stay at the University of Konstanz as Mercator 
Fellow. A.G.-M. also acknowledges funding from the Spanish MINECO (Contract No.\ MAT2014-58860-P).
\end{acknowledgments}

\appendix

\section{Convergence analysis} \label{sec-comparison}

\begin{figure}[t]
\begin{center} \includegraphics[width=0.95\columnwidth,clip]{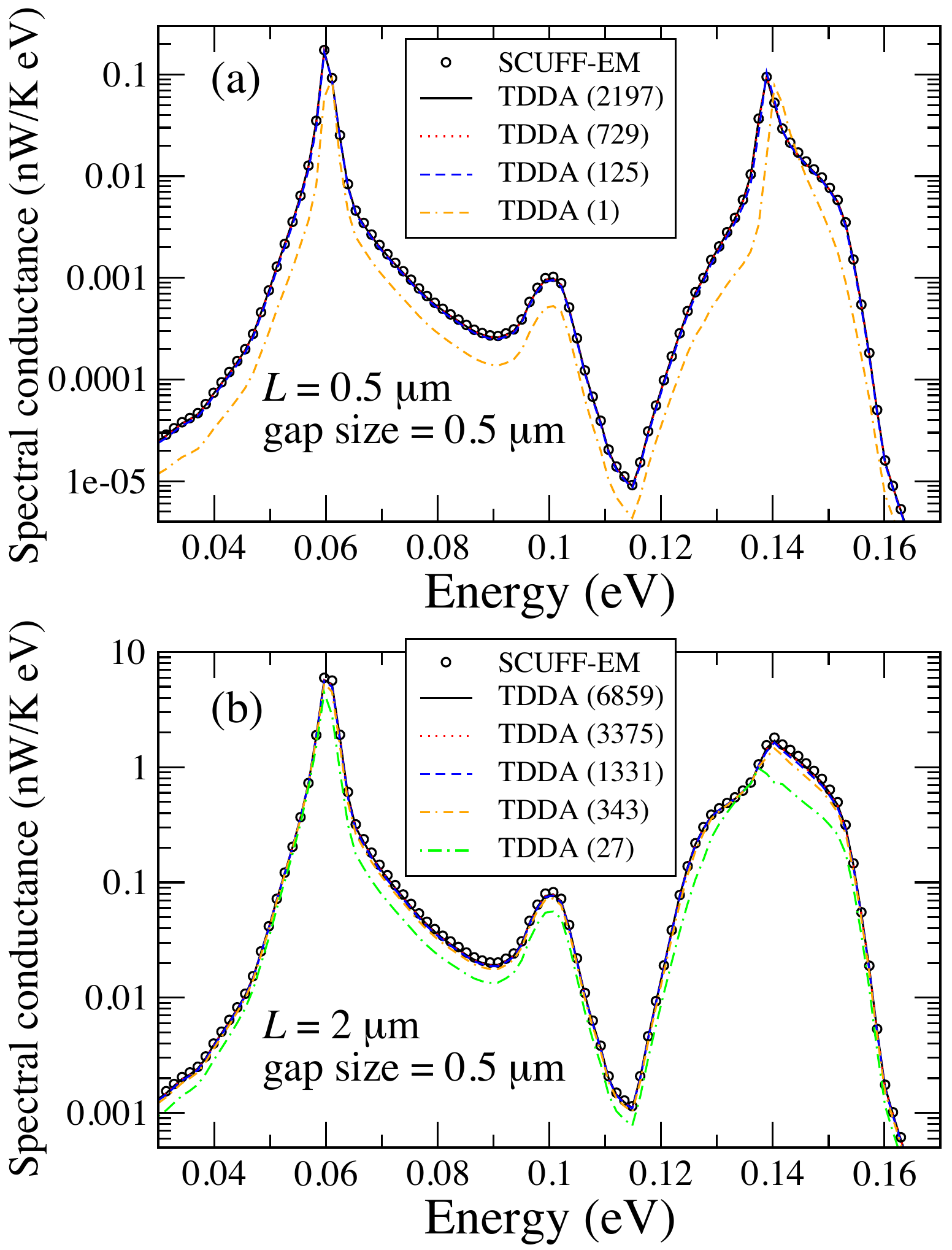} \end{center}
\caption{(Color online) (a) Spectral conductance as a function of the energy for two SiO$_2$ cubes of
side $L=0.5$ $\mu$m and a gap size of $0.5$ $\mu$m. The different lines correspond to the results computed
with the TDDA approach with different number of dipoles per cube, as indicated in the legend. The open 
symbols correspond to the converged result obtained with SCUFF-EM. (b) The same as in panel (a), but
for two SiO$_2$ cubes of side $L=2$ $\mu$m.}
\label{fig-convergence}
\end{figure}

In this appendix we present a brief analysis of the convergence of the results shown in section 
\ref{sec-results-RHT}. In Fig.~\ref{fig-convergence} we show some convergence tests for the radiative 
heat transfer between two SiO$_2$ cubes of two different sizes. In this figure, together with the 
converged results obtained with SCUFF-EM, we show the TDDA results obtained with different
number of dipoles (per cube). As one can see, upon increasing the number of dipoles, the
TDDA results first converge rapidly to approximately the correct result and then, the 
convergence improves slowly, but monotonically until a very satisfactory agreement with the SCUFF-EM
results is achieved. Of course, as the size of the objects increases, the required number of
dipoles increases accordingly and a proper convergence becomes much more demanding. The same
occurs when the gap size is reduced (not shown here). In that case, the electric field inside
the SiO$_2$ particles varies more rapidly in space due to the smaller penetration depth of the
surface phonon polaritons that dominate the NFRHT in this polar dielectric \cite{Song2015a}. All
these qualitative conclusions on the convergence trends also apply to the case of MO objects, like
that considered in Fig.~\ref{fig-RHT_2CubesInSb}. Let us conclude by reminding that a very thorough 
analysis of the convergence of TDDA for isotropic objects was reported in Ref.~[\onlinecite{Edalatpour2015}].   

\section{Alternative derivation Eq.~(\ref{eq-Pnet-bodies})} \label{sec-derivation}

For completeness, we provide here an alternative derivation of the central result of 
Eq.~(\ref{eq-Pnet-bodies}). For this purpose, we have made use of the formalism put forward by Messina
\emph{et al.}\ \cite{Messina2013} for the description of heat transfer in systems of multiple
dipoles (see also Ref.~[\onlinecite{Ben-Abdallah2011}]). This formalism was developed to study many-body 
effects in the heat transfer between dipoles, but as we show below, it can be adapted in the TDDA spirit 
to describe the heat transfer between bodies of arbitrary size.

As in section \ref{sec-TDDA}, we consider two bodies described by a collection of 
$N_1$ (body 1) and $N_2$ (body 2) electrical point dipoles. Again, each  dipole has a  
volumen $V_{i,b}$ and it is characterized by a polarizability tensor $\hat \alpha_{i,b}$, 
where $b=1,2$. In this case, rather than separating the problem into two problems of the 
emission of one body and the absorption of the other, as we did in section \ref{sec-TDDA},
we now consider that the fluctuating dipoles in both bodies are radiating at the same
time. In this case, the total dipole moments can be grouped into column supervectors 
as follows
\begin{equation}
\bar{\mathbf P} = \left( \begin{array}{cc} \bar{\mathbf P}_1 \\ 
\bar{\mathbf P}_2 \end{array} \right); \;\;
\bar{\mathbf P} =  \bar{\mathbf P}_{\rm f} +  \bar{\mathbf P}_{\rm ind} ,
\end{equation}
where there are two contributions to the total dipole moments, one coming from
the fluctuating dipoles, $\bar{\mathbf P}_{\rm f}$, and the other is an induced
contribution arising from the interaction with the other dipoles. This second 
contribution is given by \cite{Messina2013}
\begin{equation}
%\mathbf p_{{\rm ind}, i} = k^2_0 \alpha_i \sum_{j \neq i} \hat G_{ij} \mathbf p_j .
\bar{\mathbf P}_{\rm ind} = k^2_0 \bar \alpha \Delta \bar G \bar{\mathbf P} .
\end{equation}
Here, the indexes $i$ and $j$ refer to dipoles in both bodies. From this equation 
it is easy to show that the total dipole moments are given by
\begin{equation}
\label{eq-T-alt}
\bar{\mathbf P} = \bar T^{-1} \bar{\mathbf P}_{\rm f}, \; \mbox{where} \;\;
\bar T = \bar 1 - k^2_0 \bar \alpha \Delta \bar G .
\end{equation}
Here, we have used the notation of section \ref{sec-TDDA}. 

Now the goal is compute the net power balance in, let us say, body 2. 
This net power is given by 
\begin{eqnarray}     
P_{\rm net} & = & \langle \frac{d\bar{\mathbf P}_2(t)}{dt} \cdot \bar{\mathbf E}_2(t) \rangle = \\
& & \hspace*{-1cm} 2 \int^{\infty}_0 \frac{d\omega}{2\pi} \omega \int^{\infty}_{-\infty} 
\frac{d\omega^{\prime}}{2\pi} \mbox{Im} \left\{ \langle \bar{\mathbf P}_2(\omega) \cdot \bar{\mathbf 
E}^{\ast}_2(\omega^{\prime}) \rangle e^{-i (\omega - \omega^{\prime}) t} \right\} \nonumber . 
\end{eqnarray}
The dipole moment $\bar{\mathbf P}_2$ can be obtained from Eq.~(\ref{eq-T-alt}):
\begin{equation}
\bar{\mathbf P}_2 = \bar T^{-1}_{21} \bar{\mathbf P}_{{\rm f},1} + 
\bar T^{-1}_{22} \bar{\mathbf P}_{{\rm f},2} ,
\end{equation}
while $\bar{\mathbf E}_2$ can be related to $\bar{\mathbf P}_2$ via Eq.~(\ref{eq-En-pn}).
Then, with the usual algebraic manipulations, one can show that
\begin{eqnarray}
\langle \bar{\mathbf P}_2 \cdot \bar{\mathbf E}^{\ast}_2 \rangle & = & 
\frac{1}{\epsilon_0} \mbox{Tr} \left\{\bar \alpha^{-1}_2 \langle \bar{\mathbf P}_2
\bar{\mathbf P}^{\dagger}_2 \rangle \bar \alpha^{-1 \dagger}_2 \bar \chi_2 \right\} = \\
& & \frac{1}{\epsilon_0} \mbox{Tr} \left\{ \bar \alpha^{-1}_2 \bar T^{-1}_{21} 
\langle \bar{\mathbf P}_{{\rm f},1} \bar{\mathbf P}^{\dagger}_{{\rm f},1} \rangle
\bar T^{-1 \dagger}_{21} \bar \alpha^{-1 \dagger}_2 \bar \chi_2 \right. + \nonumber \\
& & \hspace*{1cm} \left. \bar \alpha^{-1}_2 \bar T^{-1}_{22} 
\langle \bar{\mathbf P}_{{\rm f},2} \bar{\mathbf P}^{\dagger}_{{\rm f},2} \rangle
\bar T^{-1 \dagger}_{22} \bar \alpha^{-1 \dagger}_2 \bar \chi_2 \right\} \nonumber ,
\end{eqnarray}
which with the help of the FDT theorem leads to
\begin{eqnarray}
P_2 & = & 2 \int^{\infty}_0 \frac{d\omega}{2\pi} \hbar \omega \times \\
& & \left[ (1+2n_{\rm B}(\omega,T_1)) \mbox{Tr} \left\{ \bar \alpha^{-1}_2 \bar T^{-1}_{21} 
\bar \chi_1 \bar T^{-1 \dagger}_{21} \bar \alpha^{-1 \dagger}_2 \bar \chi_2 \right\} + 
\right. \nonumber \\
& & \hspace*{1mm} \left. (1+2n_{\rm B}(\omega,T_2)) \mbox{Tr} \left\{ \bar \alpha^{-1}_2 \bar T^{-1}_{22} 
\bar \chi_2 \bar T^{-1 \dagger}_{22} \bar \alpha^{-1 \dagger}_2 \bar \chi_2 \right\} \right] \nonumber .
\end{eqnarray}

In thermal equilibrium ($T_1=T_2$), $P_2$ should vanish. Therefore, they following relation 
must hold
\begin{eqnarray}
\mbox{Tr} \left\{ \bar \alpha^{-1}_2 \bar T^{-1}_{21} \bar \chi_1 \bar T^{-1 \dagger}_{21} 
\bar \alpha^{-1 \dagger}_2 \bar \chi_2 \right\} & = & \\ - \mbox{Tr} \left\{ \bar \alpha^{-1}_2 
\bar T^{-1}_{22} \bar \chi_2 \bar T^{-1 \dagger}_{22} \bar \alpha^{-1 \dagger}_2 \bar \chi_2 \right\} .
& & \nonumber
\end{eqnarray}

Thus, $P_2$ can be rewritten as
\begin{equation}
P_{\rm net} = \int^{\infty}_0 \frac{d\omega}{2\pi} 
[\Theta (\omega,T_1) - \Theta (\omega,T_2)] {\cal T}(\omega) ,
\end{equation}
where
\begin{equation}
\label{eq-T-new}
{\cal T}(\omega) = 4 \mbox{Tr} \left\{ \bar \alpha^{-1}_2 \bar T^{-1}_{21} 
\bar \chi_1 \bar T^{-1 \dagger}_{21} \bar \alpha^{-1 \dagger}_2 \bar \chi_2 \right\} .
\end{equation}
Now, the remaining task is to show that the previous expression for the transmission
reduces to Eq.~(\ref{eq-T-bodies}). For this purpose, one can use Eq.~(\ref{eq-T-alt})
to show that
\begin{eqnarray}
\bar T^{-1}_{22} & = & \left[ \bar 1 - k^2_0 \bar \alpha_2 \Delta \bar G_{22} - \right. \\ & &
\left. k^4_0 \bar \alpha_2 \Delta \bar G_{21} \left[ \bar 1 - k^2_0 \bar \alpha_1 
\Delta \bar G_{11} \right]^{-1} \bar \alpha_1 \Delta \bar G_{12} \right]^{-1} , \nonumber \\
\bar T^{-1}_{21} & = & k^2_0 \bar T^{-1}_{22} \bar \alpha_2 \Delta \bar G_{21} 
\left[ \bar 1 - k^2_0 \bar \alpha_1 \Delta \bar G_{11} \right]^{-1} .
\end{eqnarray}

From these equations it is straightforward to show that
\begin{eqnarray}
\bar \alpha^{-1}_2 \bar T^{-1}_{22} & = & \bar D_{22} \bar \alpha^{-1}_2 \\
\bar \alpha^{-1}_2 \bar T^{-1}_{21} & = & k^2_0 \bar C_{21} ,
\end{eqnarray}
from which it is obvious that Eq.~(\ref{eq-T-new}) reduces to Eq.~(\ref{eq-T-bodies}).
This completes our alternative derivation of Eq.~(\ref{eq-Pnet-bodies}), which confirms
the validity of our result.

%%%%%%%%%%%%%%%%%%%%%%%%%%%%%%%%%%%%%%%%%%%%%%%%%%%%%%%%%%%%%%%%%%%%%%%%%%%%%%%%%%%%%%%%%%


\begin{thebibliography}{00}
\bibitem{note1}
Let us remark that we do not consider in this work the interaction with an 
external thermal bath. Therefore, the only radiative heat transfer occurs
between the particles.
\bibitem{note2}
Here, we have ignored the 1 appearing next to the Bose function in Eq.~(\ref{eq-FDT-Eb})
because it is cancelled by the corresponding contribution coming from vacuum.
\end{thebibliography}
\end{document}